# From Qubits to Opinions: Operator and Error Syndrome Measurement in Quantum-Inspired Social Simulations on Transversal Gates

Yasuko Kawahata †

Faculty of Sociology, Department of Media Sociology, Rikkyo University, 3-34-1 Nishi-Ikebukuro,Toshima-ku, Tokyo, 171-8501, JAPAN.

ykawahata@rikkyo.ac.jp,kawahata.lab3@damp.tottori-u.ac.jp

**Abstract:** This paper delves into the history and integration of quantum theory into areas such as opinion dynamics, decision theory, and game theory, offering a novel framework for social simulations. It introduces a quantum perspective for analyzing information transfer and decision-making complexity within social systems, employing a toric code-based method for error discrimination.Central to this research is the use of toric codes, originally for quantum error correction, to detect and correct errors in social simulations, representing uncertainty in opinion formation and decision-making processes. Operator and error syndrome measurement, vital in quantum computation, help identify and analyze errors and uncertainty in social simulations. The paper also discusses fault-tolerant computation employing transversal gates, which protect against errors during quantum computation. In social simulations, transversal gates model protection from external interference and misinformation, enhancing the fidelity of decision-making and strategy formation processes.

## 1. Introduction

This paper explores the history of research in quantum opinion dynamics, decision theory, and game theory, and how these theories form a new paradigm for social simulation. In particular, we present a new approach to analyze the complexity of information transfer and decision-making processes within social systems from a quantum theoretical perspective, using a toric code-based error discrimination methodology. Within the framework of quantum opinion dynamics, the formation and change of individual opinions are modeled by quantum theory. This makes it possible to represent phenomena such as opinion superposition and uncertainty, which cannot be captured by conventional models. In decision theory, on the other hand, the incorporation of quantum theory provides a more realistic analysis of how individual decisions affect group decisions. Furthermore, in game theory, quantum game theory will be used to explore strategic interactions among participants from perspectives that cannot be explained by conventional theories.

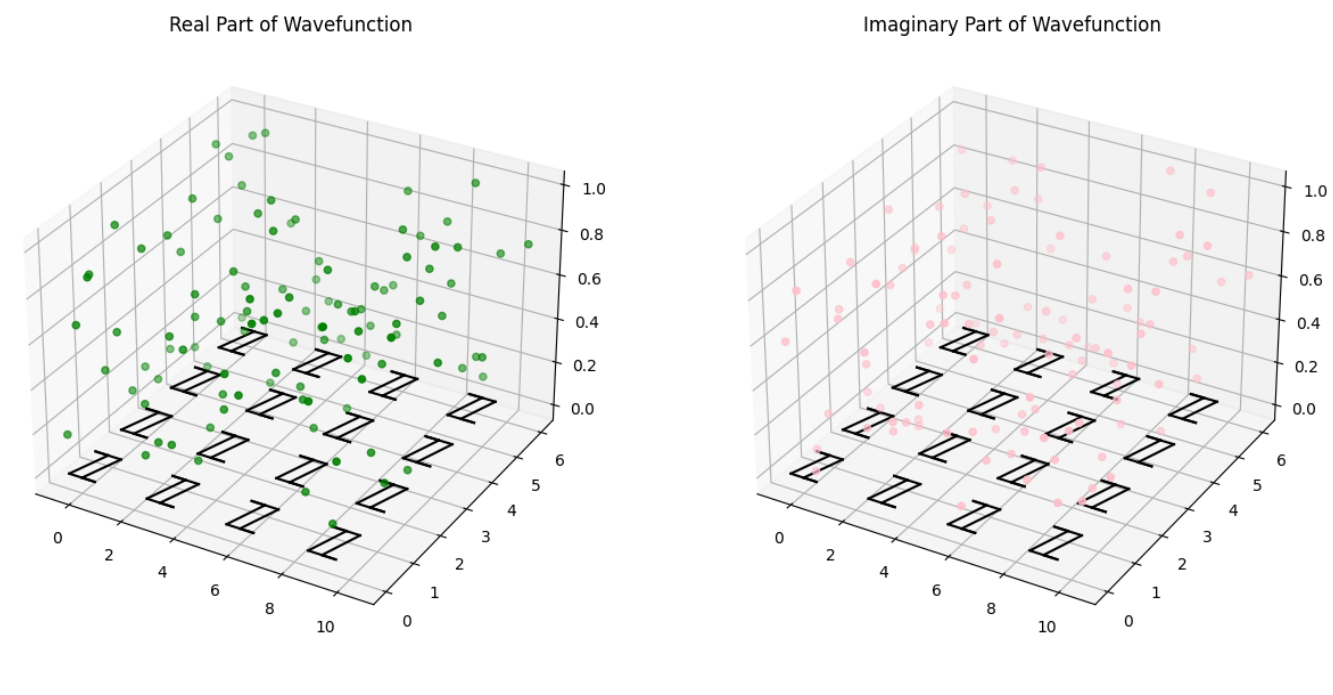

Fig. 1: Toric Code Local Opinion Distribution:$honeycomb_lattice$

At the core of this research is the concept of error discrimination using toric codes. Toric codes, developed as an error correction technique in quantum computation, are used to effectively detect and correct errors in qubits. In this study, we apply this technique to the context of social simulation and attempt to model "errors" or uncertainty in opinion formation and decision-making processes.

Also important are the concepts of operator and error syndrome measurement. In quantum computation, operators are used to change the state of qubits, and error syndrome measurement serves as a process to identify the presence and type of error. Applying this process to social simulations makes it possible to identify and analyze errors and uncertainty in opinion formation and decision processes.

Finally, the application of fault-tolerant computation using transversal gates will also be discussed. Transversal gates play an important role in quantum computation while protecting against errors during computation. In the context of social simulation, they can be used to model how decision-making and strategy formation processes are protected from external interference and misinformation. Transversal gates play an important role in quantum error correction. Using these gates, quantum computation can be performed while protecting against errors during computation. In social simulations, they can help model how decision-making and strategy formation processes are protected from external interference and misinformation.

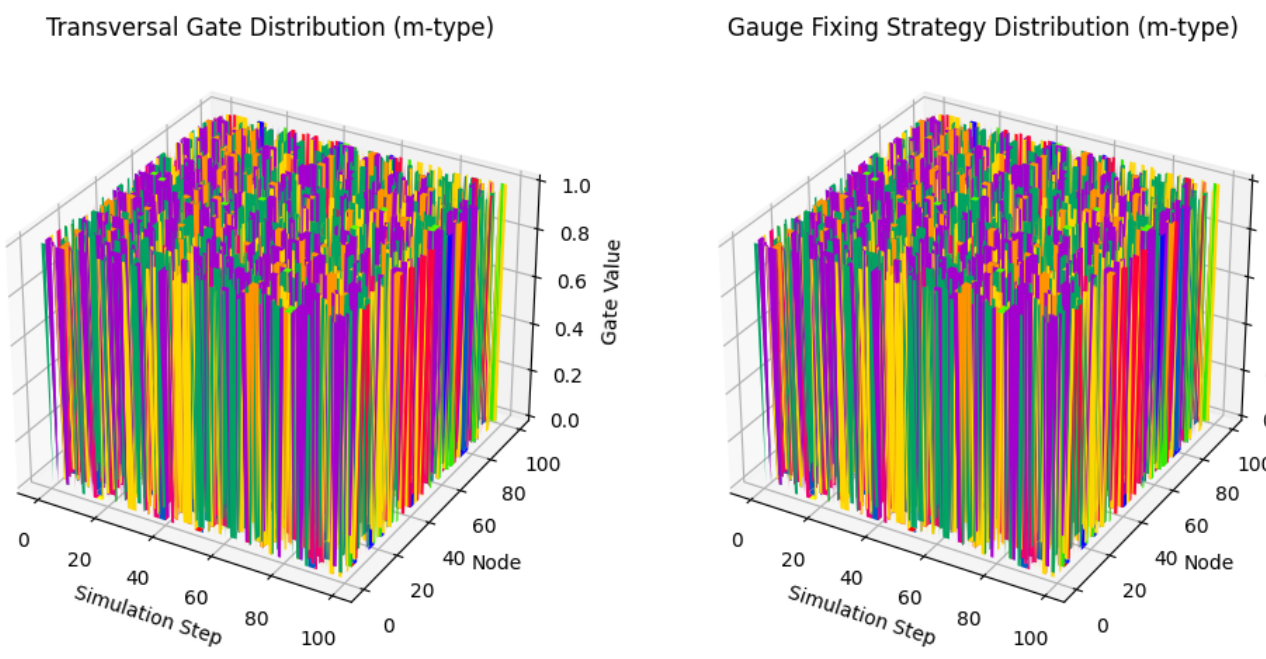

Fig. 2: Transversal Gate, Gauge Fixing Strategy Distribution

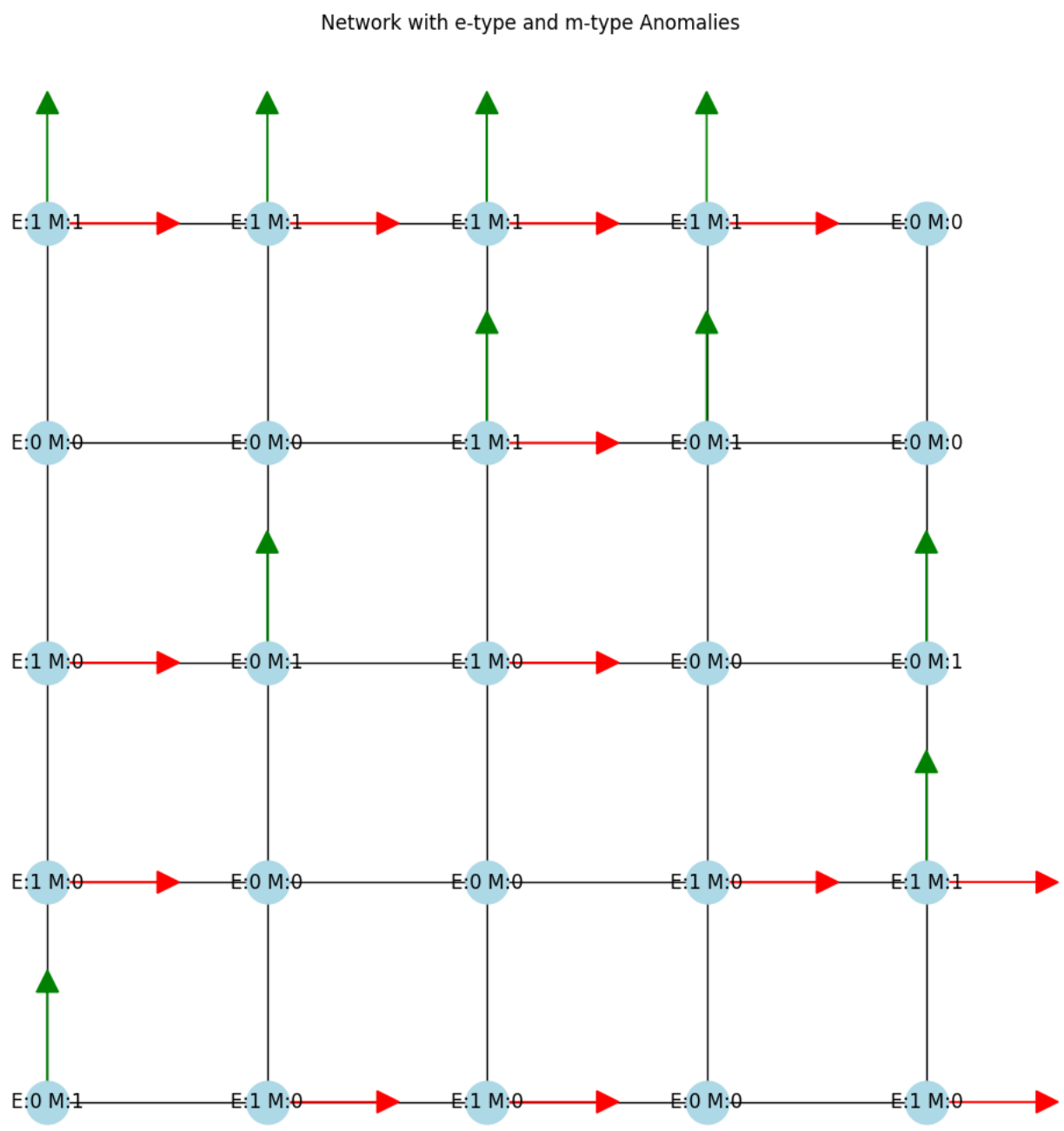

Fig. 3: Network with e-type and m-type Anomalies

Through this paper, we hope to show that quantum theory is a powerful tool for understanding and analyzing various aspects of social science from new perspectives, and to suggest prospects for research approaches that use quantum theory to analyze and gain a deeper understanding of the complexity of opinion formation and decision making in social systems.

# 2. Preview Research

## 2.1 Opinion Dynamics Research Case Studies

Much groundbreaking research has been conducted in the field of opinion dynamics. For example, in "Mixing beliefs among interacting agents" by Deffuant et al. (2000), a simple model was proposed to simulate the dynamics of opinion convergence and polarization. The model is based on the assumption that agents exchange opinions with each other and are influenced by each other if the difference in opinion is below a certain threshold. On the other hand, Hegselmann and Krause (2002) in their study "Opinion dynamics and bounded confidence: models, analysis and simulation" introduced the so-called "Bounded Confidence Model" that takes into account the influence of individual beliefs and uncertainty on the convergence of opinions. Bounded Confidence Model" was introduced. In their model, agents consider only the opinions of other agents whose opinions are somewhat similar to their own, and this leads to polarization of opinions and the formation of multiple opinion clusters. These studies revealed the complexity of interactions between individuals in the process of social opinion formation and made important contributions to our understanding of opinion dynamics. In sum, the field of opinion dynamics is comprised of diverse approaches that seek to elucidate the various factors behind the formation and change of opinions through the simulation of social interactions.

## 2.2 Research Case Study on Quantum Adamar Computation

Research on quantum adamar computation has made important contributions as a way to extend the efficiency and capabilities of quantum information processing. One of the leading works in this area is "Fault-Tolerant Quantum Computation with Constant Error Rate" by Aharonov and Ben-Or (1997). Aharonov and Ben-Or's work laid the foundation for reducing the impact of errors in quantum computation and achieving more reliable computation. and more reliable computation. In addition, Silvano Garnerone, Paolo Zanardi, and Daniel A. Lidar's (2012) "Adiabatic Quantum Algorithm for Network Community Detection" utilizes quantum adamantine computation to An algorithm was developed to efficiently detect community structure within a network, demonstrating the superior capabilities of quantum computation in analyzing complex networks. These studies clearly demonstrate that quantum Adamar computation plays an im-portant role in a wide variety of applications of quantum information processing and open new avenues of research in the field of quantum computation. Overall, these studies on quantum adamantine computation have made important contributions in the evolution of quantum computation techniques and have laid the groundwork for further developments in this field.

## 2.3 Research Case Studies on Toric Codes

Research on toric codes represents an important advance in quantum error correction and quantum information theory. For example, Alexey Kitaev's (1997) landmark paper, "Fault-tolerant quantum computation by anyons," established the basic principles of quantum error correction and quantum computation using toric codes. This work by Kitaev showed that the toric code is naturally tolerant to errors and how this helps in the stable storage and manipulation of quantum information. A subsequent paper by H. Bombin and M.A. Martin-Delgado (2006), "Topological Quantum Distillation," extended the concept of the toric code and proposed methods to further improve the efficiency of quantum error correction. Their work is characterized by the ability to correct more complex error patterns by exploiting the redundancy of quantum information. These studies show how toric codes play a fundamental role in quantum information science and provide important guidance in the design and implementation of quantum computers. Overall, these studies on toric codes promote a deeper understanding of quantum error correction in theory and practice, and represent important advances in the field of quantum computing.

## 2.4 Research on Applications of Toric Codes Case Study

Research on applications of toric codes has made important contributions in the areas of quantum error correction and quantum computing. For example, in "Surface codes: Towards practical large-scale quantum computation" by Austin G. Fowler et al. (2012), the concept of surface codes based on toric codes was described in detail and their application to large-scale quantum computation Applications to large-scale quantum computation were proposed. This work provided concrete strategies for increasing the efficiency and practicality of quantum error correction and was considered an important step toward the realization of quantum computers. In addition, Benjamin J. Brown et al. (2016), "Fault-tolerant error correction with the gauge color code," introduced a new class of quantum error-correcting codes called gauge color codes, which are toric codes and showed that it can serve as a generalization of the Their work further extended the concept of the toric code and opened up new possibilities for quantum error correction. These studies show how the toric code plays a fundamental role in the development of quantum information theory and represent important advances in the field of quantum computing. Overall, these applied studies on the toric code promote a deeper understanding of quantum error correction in theory and practice, and lay an important foundation for the future of quantum computing.

## 2.5 Applied Research Examples of Gauge Color Codes

Research on gauge color codes represents a new development in the field of quantum error correction. In particular, in "Gauge Color Codes: Optimal Transversal Gates and Gauge Fixing in Topological Stabilizer Codes" by Bombin et al. (2013), optimal quantum error correction using gauge color codes with Transversal Gates and Gauge Fixing Strategies for Quantum Error Correction Using Gauge Color Codes was developed. This study showed that the use of gauge color codes in topological stabilizer codes can significantly improve the efficiency and effectiveness of error correction in the quantum computation process. Also, in "Universal Color-Code Quantum Computation" by Kubica et al. (2015), a framework for universal quantum computation using gauge color codes was proposed and how this code can be used to perform complex algorithms and operations in quantum computation and how this code can be used to perform complex algorithms and operations in quantum computation. These studies clearly demonstrate that gauge-color codes play an important role in the field of quantum error correction and are an important step toward the practical application of quantum computing. In summary, these applied studies on gauge color codes contribute to the further development of quantum error correction techniques and pave the way toward the realization of quantum computing. An important research case study on gauge color codes is "Gauge Color Codes: Optimal Transversal Gates and Gauge Fixing in Topological Stabilizer Codes" by Héctor Bombín (2015) Bombín's work proposes a new approach in quantum error correction and shows that the use of these codes within topological stabilizer codes can improve the efficiency and effectiveness of quantum error correction. and suggests how gauge color codes can be useful in performing complex algorithms and operations in quantum computation. The subsequent work by Aleksander Kubica et al. (2015), "Universal Color-Code Quantum Computation," proposed a framework for universal quantum computation using gauge color codes. codes play an important role not only in the field of quantum error correction, but also in the practical application of quantum computation. Taken together, these research examples make it clear that gauge color codes play a central role in the development of quantum information theory and are an essential technology for the realization of quantum computing.

## 2.6 Research Case Studies on Transversal Gates

Research on transversal gates has led to important advances in quantum computing, as Daniel Gottesman (1998) in "The Heisenberg Representation of Quantum Computers, Gottesman's work was a breakthrough in the integration of quantum error correction and quantum computation, and had a major impact on later research in quantum computing. and had a major impact on later research in quantum computing. Efficient fault-tolerant quantum computing" by Andrew Steane (1999) developed specific methods for fault-tolerant quantum computation using transversal gates, making error correction in quantum systems more realistic. Steane's work is an important step toward practical quantum computing and strengthens the theoretical framework for quantum error correction. These research examples illustrate that transversal gates play a central role in the areas of error correction and fault-tolerant computation in quantum computing. In general, these studies on transversal gates have made important contributions in the evolution of quantum computing technology and have laid the groundwork for further developments in this field.

## 2.7 Research Case Study on Fault-Tolerant Computation

The field of fault-tolerant computation has seen significant research to increase the reliability and feasibility of quantum computing; Peter Shor's (1996) "Fault-Tolerant Quantum Computation" is one of the pioneering works in this field, Shor's work addressed the vulnerabilities of qubits and led to important advances in the design of practical quantum computers. Andrew Steane's (1997) subsequent work, "Active Stabilization, Quantum Computation, and Quantum State Synthesis," explored the possibility of fault-tolerant quantum computation using active stabilization techniques and provided a quantum error-correcting codes, Steane's work showed how the theory of quantum error correction could be applied to actual quantum computation processes and paved the way for the realization of quantum computation. These research examples clearly demonstrate that fault-tolerant computation is a central approach to ensuring error correction and reliability in quantum computing. Overall, these studies on fault-tolerant computation are an integral part of the evolution of quantum computing technology and lay the groundwork for further developments in the field.

## 2.8 Research Case Studies in Quantum and Social Simulation

Research on the convergence of quantum computing and social simulation explores new boundaries between the two fields.In Quantum Social Science, Jacob Biamonte and Peter Wittek (2019) explore how quantum theory can be applied to social science problems, especially decision making, game theory, and social network analysis. The research focused on the possibilities of going beyond classical social science models to take advantage of the new perspectives and computational power offered by quantum computing principles. On the other hand, Quantum Techniques for Stochastic Mechanics by Eleanor Rieffel et al. (2018) explored how quantum algorithms can be used in social science modeling, particularly in the analysis of stochastic phenomena. Their work is credited with opening new avenues for simulating more complex social systems by taking advantage of the computational power offered by quantum computing. These research cases bridge the gap between quantum computing and the social sciences, opening up new areas of understanding and analysis of social systems. In sum, quantum and social simulation research shows promise in applying quantum computing principles to solve social science problems and lays an important foundation for future research and applications.

## 2.9 Quantum and Game Theory Research Case Study

Research on the integration of quantum computing and game theory has provided innovative insights into the field. in his paper "Quantum Strategies", David A. Meyer (1999) extended the traditional game theory concept of strategy to the quantum world and analyzed the new benefits that players can gain by exploiting quantum states. Meyer's work showed that quantum mechanics adds a new dimension to game theory and laid the foundation for quantum game theory. Subsequently, Eisert, Wilkens, and Lewenstein's (1999) "Quantum Games and Quantum Strategies" further embodied the concepts of quantum game theory by showing how quantum strategies can provide advantages in classical games such as the prisoner's dilemma. They further embodied the concepts of quantum game theory. Their work proposed the development of a new type of game theory that exploits the properties of quantum information and was an important step in the integration of the fields of quantum and game theory. From these research examples, it is clear that quantum game theory provides a new framework for understanding and predicting the outcomes of games and can add a new perspective to traditional game theory by exploiting the principles of quantum computing. Overall, these studies on the integration of quantum and game theory play an innovative role in the development of game theory and open up new research possibilities.

## 2.10 Research Cases on Quantum and Decision Making

Research on the relevance of quantum theory and decision making offers a new paradigm for decision theory. in "Quantum Models of Cognition and Decision" by Jerome R. Busemeyer and Peter D. Bruza (2012), quantum probability theory is applied to psychological It focuses on the application of quantum probability theory to the modeling of psychological decision-making processes. The study noted that human decision making can exhibit contradictions and paradoxes that cannot be explained by traditional classical probability models, and showed that quantum theory can better model these phenomena. Furthermore, in "Quantum Probability Theory in Decision Making: From Quantum Physics to Economics and Social Science" by Zheng Wang (2013), he applied decision-making problems in economics and social science to Wang's research demonstrates the potential of the concepts provided by quantum theory to extend decision theory in economics and social science. These research examples show that quantum theory offers a new perspective on modeling decision making and provides a new theoretical framework in psychology, economics, and the social sciences. Overall, the study of quantum and decision making challenges traditional theories and adds a new dimension to our understanding of decision making, opening up new research possibilities in these fields.

### 2.11 Research Case Studies in Quantum and Network Analysis

Research on the application of quantum theory to network analysis offers new methods for better understanding and analyzing networks. Notable work in this area includes "Quantum Random Walks: A New Method for Designing Quantum Algorithms" by Vittorio Giovannetti, Seth Lloyd, and Lorenzo Maccone (2008) by Seth Lloyd, Lorenzo Maccone (2008). In this study, a new type of quantum algorithm based on quantum random walks was proposed, which allows efficient exploration and analysis on networks. pioneered a new way to explore and analyze the complex structure of networks more efficiently using quantum theoretical principles. In addition, Silvano Garnerone, Paolo Zanardi, and Daniel A. Lidar's (2012) "Adiabatic Quantum Algorithm for Network Community Detection" used quantum adamantine computation to develop an algorithm for detecting community structure in networks. The work developed an algorithm for detecting community structure within a network using quantum adiabatic computation. This research demonstrated the potential of quantum computation to go beyond classical algorithms in analyzing the structure of networks, and showed new avenues for coupling network theory and quantum information science. These research examples clearly demonstrate that the integration of quantum theory and network analysis plays an important role in fostering the development of new computational methods and analytical techniques in network theory and in improving our understanding of complex networks. In sum, quantum and network analysis research is opening new frontiers in network science and expanding the possibilities for research in this field.

### 2.12 Research Case Studies in Quantum and Opinion Dynamics

Research applying quantum theory to opinion dynamics offers new perspectives in decision theory and the social sciences, and Peter D. Bruza et al.'s (2009) study "Quantum Models of Cognition as an Alternative Explanation for Some "Paradoxes" in Opinion Dynamics," attempted to explain the traditional paradoxes in opinion dynamics using quantum cognitive models. The study proposed that principles of quantum theory can better capture the processes of opinion formation and decision making, presenting a new approach that goes beyond the limitations in classical opinion dynamics models. In addition, Zheng Wang's (2013) paper, "Quantum Theory and the Dynamics of Opinion," focuses on how quantum theory can help model the dynamics of opinion evolution and change, taking advantage of the stochastic and interference properties that quantum theory offers The authors showed that quantum theory can be used to more accurately capture the formation and change of social opinions. These research examples demonstrate the potential of applying quantum theory to the social sciences and provide a new theoretical framework for opinion dynamics. Overall, these studies on the integration of quantum and opinion dynamics add a new dimension to theories of opinion formation and make important contributions to extending research methods and understanding in the social sciences.

## 3. Discussion

### 3.1 Plaquette Operator and Star Operator

The Plaquette operator and Star operator play crucial roles in the field of quantum error correction, particularly in toric codes. They are central elements in understanding the concept of topological quantum error correction.

### 3.2 Plaquette Operator

In toric codes, an operator corresponding to each cell (plaquette) of the lattice is defined. Considering a two-dimensional square lattice, each plaquette refers to one of the squares of the lattice. The Plaquette operator $B_p$ acts on the four qubits (quantum bits) associated with the edges adjoining that plaquette. Typically, this is represented using the Z basis (Pauli Z operator). It is expressed in the following formula:

$$B_p = \sigma_1^z \sigma_2^z \sigma_3^z \sigma_4^z$$

where $\sigma_i^z$ is the Pauli Z operator acting on the $i$-th qubit.

### 3.3 Star Operator

The Star operator $A_s$ is associated with each vertex of the lattice. It acts on the four edges (or qubits) that are connected to each vertex. This is usually represented using the X basis (Pauli X operator) and is defined as follows:

$$A_s = \sigma_1^x \sigma_2^x \sigma_3^x \sigma_4^x$$

where $\sigma_i^x$ is the Pauli X operator acting on the $i$-th qubit.

## 3.4 Role in Toric Codes

In toric codes, these operators are used for error detection and correction. The Plaquette operator and Star operator define how the quantum state reacts to errors, and by measuring the eigenvalues of these operators, the presence or absence of errors can be detected. Importantly, these operators are commutative; that is, the order of the operators does not affect the outcome.

Toric codes are an example of topological quantum error correction, and these operators play a central role in the construction of such error correction codes. Such codes are a crucial step toward the realization of future quantum computers.

When talking about the 'trajectory' of a quantum error correction code, it generally refers to the possible paths that a quantum state can take during the error correction process or the set of states defined by a specific error correction code. There are several important features of the trajectory of an error correction code:

(1) **Error Correction Capability**: The trajectory of an error correction code has the ability to correct a specific type and number of errors. For instance, it may be able to correct some simple errors (like bit flips, phase flips) but not more complex errors.

(2) **Code Space**: The code space corresponding to a specific error correction code is constructed based on the types and numbers of permissible errors. This space includes the basis states that represent the 'ideal' quantum states where no errors have occurred.

(3) **Error Identification**: In quantum error correction, it is important to detect whether an error has occurred and, if possible, to identify its type. The trajectory is defined through the process of error detection and identification.

(4) **Distinction Between Logical and Physical Operations**: The trajectory includes the distinction between errors (physical operations) on the physical qubits and errors (logical operations) on the encoded information.

(5) **Redundancy**: Redundancy of quantum information is essential for effective error correction. This involves encoding the same information across multiple qubits so that if some qubits experience errors, the information is not lost.

(6) **Set of Tolerable Errors**: The trajectory of a certain error correction code is optimized for a specific set of errors. This depends on the error model for which the code was designed.

(7) **Topological Features**: In topological quantum error correction codes (like the toric code), the trajectory has topological features and is robust against local changes in space.

These features are important in the design and analysis of quantum error correction codes. Error correction codes are fundamental for enhancing the feasibility of quantum computation and improving the resilience of quantum systems against errors.

## 3.5 Error Syndrome Measurement

Error syndrome measurement is used in toric codes and other quantum error correction codes to identify where an error has occurred. The specific formulas vary depending on the type of error correction code used, but here we explain it using the toric code as an example.

In toric codes, error syndrome measurement is carried out using the Plaquette operator $B_p$ and the Star operator $A_s$.

### 3.5.1 Plaquette Operator $B_p$

The Plaquette operator is associated with each plaquette (cell) and is defined as follows:

$$B_p = \sigma_1^z \sigma_2^z \sigma_3^z \sigma_4^z$$

where $\sigma_i^z$ is the Pauli Z operator, applied to the four qubits adjacent to the plaquette.

### 3.5.2 Star Operator $A_s$

The Star operator is associated with each vertex (star) and is defined as:

$$A_s = \sigma_1^x \sigma_2^x \sigma_3^x \sigma_4^x$$

where $\sigma_i^x$ is the Pauli X operator, applied to the four qubits forming the arms of the star.

Error syndrome measurement involves measuring the eigenvalues of these operators. In an ideal state, the eigenvalues of these operators are +1. However, when an error occurs, the eigenvalues may become -1.

- If the eigenvalue of the Plaquette operator $B_p$ is -1, it indicates a Z-direction error (phase flip) in the qubits adjacent to that plaquette. - If the eigenvalue of the Star operator $A_s$ is -1, it indicates an X-direction error (bit flip) in the qubits of that star.

To mathematically represent situations in toric codes where the eigenvalues of the Plaquette operator $B_p$ and the Star operator $A_s$ become -1, basic knowledge of Pauli operators and the state of qubits is required. Below, we specifically describe how these operators detect errors.

### 3.6 Case of Plaquette Operator $B_p$

The Plaquette operator $B_p$ is defined as:

$$B_p = \sigma_1^z \sigma_2^z \sigma_3^z \sigma_4^z$$

where each $\sigma_i^z$ is the Pauli Z operator. These operators have the effect of flipping the phase of a qubit but do not change the state itself when the qubit is in $|0\rangle$ or $|1\rangle$ state.

- **No error**: If all qubits are error-free, the eigenvalue of $B_p$ is +1. That is, applying $B_p$ does not change the overall state of the system. - **With a Z-direction error**: If one or more qubits adjacent to the plaquette have a Z-direction error (application of $\sigma^z$), the eigenvalue of $B_p$ becomes -1. In this case, applying $B_p$ inverts the overall state of the system.

### 3.7 Case of Star Operator $A_s$

The Star operator $A_s$ is defined as:

$$A_s = \sigma_1^x \sigma_2^x \sigma_3^x \sigma_4^x$$

where each $\sigma_i^x$ is the Pauli X operator. These operators have the effect of flipping the state of a qubit from $|0\rangle$ to $|1\rangle$ or vice versa.

- **No error**: If all qubits are error-free, the eigenvalue of $A_s$ is +1. That is, applying $A_s$ does not change the overall state of the system. - **With an X-direction error**: If one of the four qubits forming the star has an X-direction error (application of $\sigma^x$), the eigenvalue of $A_s$ becomes -1. In this case, applying $A_s$ inverts the overall state of the system.

Let's examine in detail the formulas and computational processes for error syndrome measurement and error correction processes for a single bit flip error in toric codes. For simplicity, we use the example of a small toric code with four qubits.

### 3.8 Initial State

Let's assume that the initial state is where all qubits are in the $|0\rangle$ state. That is, the state of the system is $|0000\rangle$.

### 3.9 Occurrence of Error

Suppose a bit flip error (application of the Pauli X operator $\sigma^x$) occurs on one qubit. If this error occurs on the first qubit, the state of the system changes as follows:

$$\sigma_1^x |0000\rangle = |1000\rangle$$

### 3.10 Error Syndrome Measurement

In toric codes, error syndrome is measured using the Star operator $A_s$ and the Plaquette operator $B_p$. Here, we focus on the Star operator.

The Star operator $A_s$ is defined as follows (for simplicity, consider only one vertex):

$$A_s = \sigma_1^x \sigma_2^x \sigma_3^x \sigma_4^x$$

Applying this operator to the error state $|1000\rangle$ gives:

$$A_s|1000\rangle = (\sigma_1^x \sigma_2^x \sigma_3^x \sigma_4^x)|1000\rangle$$

Using the properties of the Pauli X operator $\sigma^x|0\rangle = |1\rangle$ and $\sigma^x|1\rangle = |0\rangle$, we get:

$$= (\sigma_1^x|1\rangle)(\sigma_2^x|0\rangle)(\sigma_3^x|0\rangle)(\sigma_4^x|0\rangle)$$

$$= |0000\rangle$$

This result indicates that the eigenvalue of the Star operator $A_s$ is -1, as the original state $|1000\rangle$ has changed.

## 4. Error Correction

The error syndrome indicates that there is an error in the first qubit. To correct this error, we again apply a bit flip ($\sigma^x$):

$$\sigma_1^x|1000\rangle = |0000\rangle$$

This returns the system to its original error-free state $|0000\rangle$.

Using the Star operator, we have measured the error syndrome for a bit flip error and performed the appropriate error correction operation (applying the bit flip again to the same qubit), returning the system to its original error-free state. This process demonstrates how the toric code can efficiently detect and correct a single bit flip error.

To understand the computational process of the error syndrome for the Star operator $A_s$ in toric codes, it is necessary to understand the action of the Pauli X operator and its eigenvalues.

### 4.1 Action of the Pauli X Operator $\sigma^x$

The Pauli X operator $\sigma^x$ has the effect of inverting the state of a qubit. That is, - $\sigma^x|0\rangle = |1\rangle$ - $\sigma^x|1\rangle = |0\rangle$

### 4.2 Action of the Star Operator $A_s$

The Star operator $A_s$ is the product of four Pauli X operators.

$$A_s = \sigma_1^x \sigma_2^x \sigma_3^x \sigma_4^x$$

#### 4.2.1 No Error Case

In a state where all qubits are error-free (that is, in a state that is an eigenstate of $\sigma^x$), applying $A_s$ does not change the overall state of the system. This is because applying $\sigma^x$ to each qubit inverts their state, but overall, the system returns to its original state. In this case, the eigenvalue of $A_s$ is +1.

#### 4.2.2 With an X-direction Error

If one of the four qubits forming the star has an X-direction error (for example, the state is inverted by $\sigma^x$), the eigenvalue of $A_s$ becomes -1. This is because when $\sigma^x$ is applied to the qubit that is inverted by the error, it returns to its original state. As a result, the overall state of the system is inverted, and the eigenvalue of $A_s$ becomes -1.

## 4.3 Mathematical Representation

Mathematically, the action of $A_s$ in the case of no error and with an X-direction error can be represented as follows: - **No Error Case**: $A_s|\rangle = |\rangle$ (where $|\rangle$ is any error-free state) - **With an X-direction Error**: $A_s|'\rangle = -|'\rangle$ (where $|'\rangle$ is a state containing one or more X-direction errors)

To understand the computational process of the error syndrome for the Plaquette operator $B_p$ in toric codes, it is first necessary to know about the action of the Pauli Z operator and its eigenvalues.

## 4.4 Mathematical Representation

Mathematically, the action of $A_s$ can be expressed as follows for the cases with and without an error:

**No Error**: $A_s|\rangle = |\rangle$ (where $|\rangle$ is any state without error).

**X-Direction Error Present**: $A_s|'\rangle = -|'\rangle$ (where $|'\rangle$ is a state with one or more X-direction errors).

To understand the calculation process of the error syndrome for the Plaquette operator $B_p$ in toric codes, it is first necessary to understand the action and eigenvalues of the Pauli Z operator.

## 4.5 Action of the Pauli Z Operator $\sigma^z$

The Pauli Z operator $\sigma^z$ acts as follows:

$$\sigma^z|0\rangle = |0\rangle$$

$$\sigma^z|1\rangle = -|1\rangle$$

That is, $\sigma^z$ does nothing to the state $|0\rangle$ and applies a coefficient of -1 to the state $|1\rangle$.

## 4.6 Action of the Plaquette Operator $B_p$

The Plaquette operator $B_p$ is the product of four Pauli Z operators:

$$B_p = \sigma_1^z\sigma_2^z\sigma_3^z\sigma_4^z$$

#### 4.6.1 No Error

When all qubits are in their eigenstates of $\sigma^z$, such as the $|0\rangle$ or $|1\rangle$ state, applying $B_p$ applies $\sigma^z$ to each qubit. If all qubits are error-free (i.e., in the eigenstate of $\sigma^z$), the eigenvalue of $B_p$ is +1. This is because each qubit gets a coefficient of +1 or -1 from $\sigma^z$, and the total product results in +1.

#### 4.6.2 Z-Direction Error Present

If one or more qubits experience a Z-direction error (for example, an error that changes $|0\rangle$ to $|1\rangle$), the action of $\sigma^z$ on those qubits does not change the state but applies a coefficient of -1. If an odd number of qubits experience such an error, the eigenvalue of $B_p$ becomes -1. This is because an odd number of -1 coefficients results in a total product of -1.

## 4.7 Mathematical Expression

Mathematically, the action of $B_p$ can be expressed as follows for the cases with and without an error:

**No Error**: $B_p|\rangle = |\rangle$ (where $|\rangle$ is any state without error).

**Z-Direction Error Present**: $B_p|'\rangle = -|'\rangle$ (where $|'\rangle$ is a state with one or more Z-direction errors).

**Network Nodes**: Each node in the network represents individuals, communities, or organizations within a social system.

**Node State**: The state of each node is represented by a numerical value indicating the node's health, economic status, social influence, etc. A normal state is denoted as '0', and an abnormal state as '1'.

## 4.8 Error Syndrome Measurement

**Detection of Anomalies**: A sudden change in the state of a node to '1' is detected as a bit-flip error (social or economic anomaly).

**Assessment of Impact Range**: The impact of anomalies originating from a node is assessed on surrounding nodes using the adjacency matrix to consider the connections between nodes.

## 4.9 Error Correction Process

**Local Correction**: Direct interventions (such as economic or social support) are implemented for nodes where anomalies are detected.

**Global Correction**: If the anomaly affects the entire system, a broader policy review and system-level interventions become necessary.

**Detection of Anomalies**:

$$E_i = \begin{cases} 1 & \text{if node } i \text{ is in an abnormal state} \\ 0 & \text{otherwise} \end{cases}$$

**Assessment of Impact Range**:

$$I_i = \sum_{j=1}^{N} A_{ij}E_j$$

Here, $I_i$ represents the total abnormal impact on node $i$, and $A_{ij}$ is the element of the adjacency matrix indicating the connection between nodes $i$ and $j$.

(1) **Anomaly Detection**: Evaluate the state of each node and identify abnormal states (bit-flip errors).

(2) **Calculation of Impact Range**: Calculate the impact range of the nodes where anomalies have occurred.

(3) **Decision of Corrective Actions**: Based on the detected anomalies and their impact range, determine appropriate interventions.

We propose formulas and calculation processes for applying the concepts of 'charge' (e-type) anyons and 'magnetic flux' (m-type) anyons from toric codes to a social network model. In toric codes, e-type anyons are detected by the Star operator $A_s$, and m-type anyons by the Plaquette operator $B_p$. Applying these to social systems corresponds to modeling different types of 'anomalies' and 'impacts'.

#### 4.9.1 1. e-type anyons (Charge-Type Anomalies)

**Modeling**: e-type anyons can be considered as representing individual 'local anomalies' within the social system, such as sudden economic crises or social unrest in specific areas.

**Formulas**:

$$E_i = \begin{cases} 1 & \text{if node } i \text{ has a local anomaly} \\ 0 & \text{otherwise} \end{cases}$$

**Calculation Process**: Evaluate the state of each node and detect local anomalies.

#### 4.9.2 2. m-type anyons (Magnetic Flux-Type Anomalies)

**Modeling**: m-type anyons represent more extensive 'global anomalies' or imbalances across the entire system, such as overall economic instability or widespread social dissatisfaction.

**Formulas**:

$$M = \sum_{i=1}^{N} E_i$$

Here, $M$ represents the total anomaly across the system.

**Calculation Process**: Aggregate the number of anomalies across the entire system and evaluate the overall imbalance.

**Local Interventions**: Plan interventions of appropriate scale and intensity based on the energy of local anomalies.

**Global Strategies**: Consider more extensive policy changes and system-level interventions based on the energy of global anomalies.

This approach allows for more effective identification of different types of problems and optimal responses to each. By using the concept of band structure, the 'weight' and 'impact' of problems can be understood in more detail, allowing for the prioritization of interventions based on these aspects.

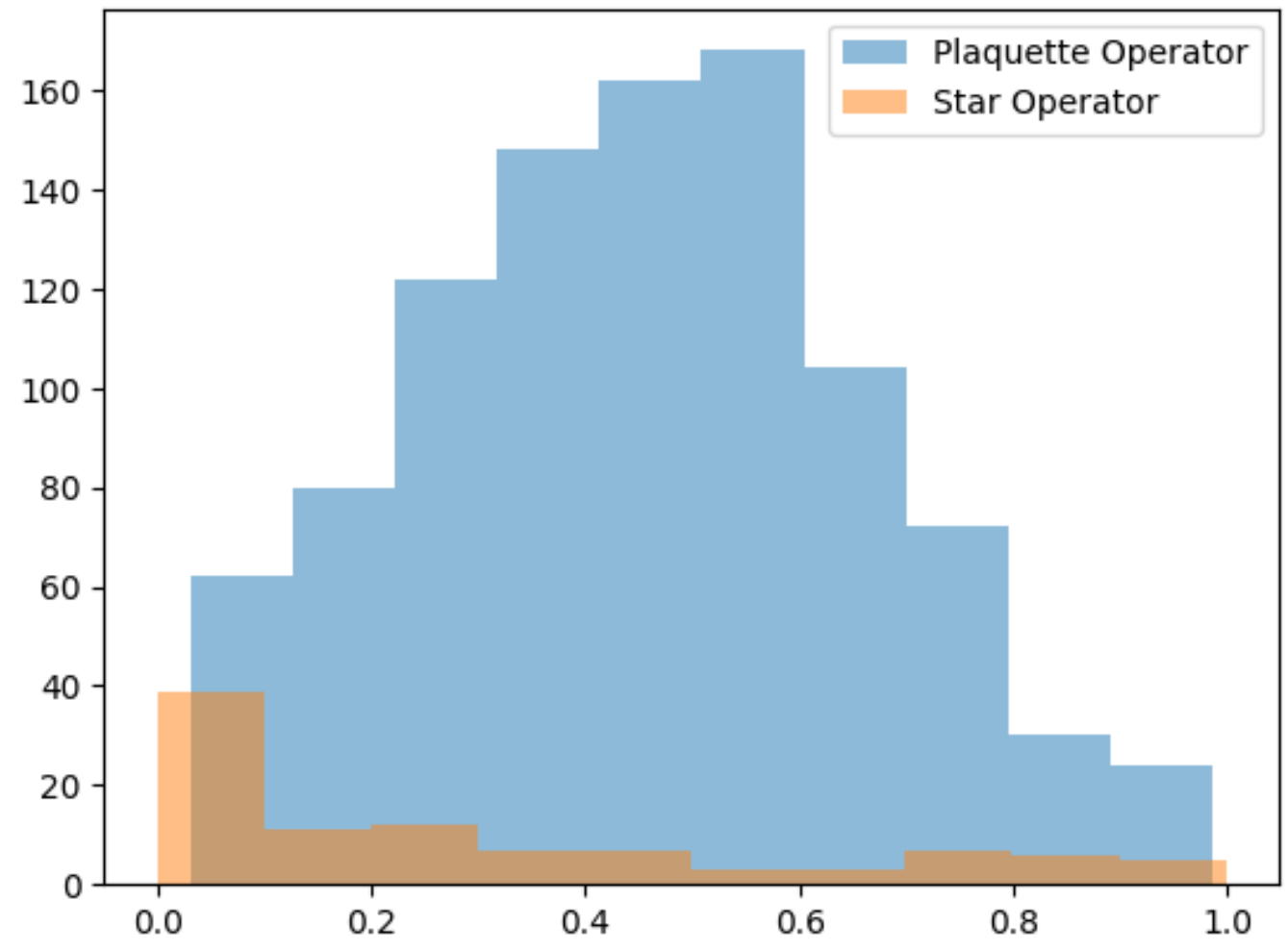

Fig. 4: Two different kinds of operators or states in the network

Results of Toric Code model, given the mention of plaquette and star operators, which are concepts from that model. In the context of opinion dynamics, the nodes of the network could represent individuals, and their states could represent their opinions on a certain topic. The update rules defined by the plaquette and star operators would then correspond to the way these opinions change due to interactions with other individuals in the network. For example, the plaquette operator might represent the consensus of opinion between neighboring individuals, while the star operator might represent the self-reinforcement or individual's tendency to become more extreme in their opinion.

## Interaction Related to the Star Operator ($J_s$)

- In quantum error correction, the star operator typically involves a product of Pauli matrices around a given vertex (or node) in the lattice, affecting its state. In your code, 'g(states)' could be analogous to the star operator, with 'states**2' indicating that the interaction strength or effect increases with the square of the state value. This could mean that the more 'opinionated' or 'extreme' a node is (higher state value), the stronger the effect of this operator.

## Interaction Related to the Plaquette Operator ($J_p$)

The plaquette operator usually involves a product of Pauli matrices around a face (or plaquette) in the lattice, which affects the states around that plaquette. In opinion dynamics, this could be interpreted as a way to reach a middle ground or average opinion among connected individuals. The function '(x + y) / 2' could represent a simple averaging of states, suggesting a smoothing or consensus-forming interaction.

## Term Related to Charge (e-type) Anyons ($H_e$)

- In the Toric Code and related models, e-type anyons are associated with errors on qubits that the plaquette operator can detect. In opinion dynamics, this might be related to discrepancies or 'errors' in the consensus process, where the plaquette operator could reveal nodes whose opinions are out of sync with their neighbors.

## Term Related to Flux (m-type) Anyons ($H_m$)

- Similarly, m-type anyons are associated with errors that the star operator can detect. In the context of your model, this could relate to individual nodes that are becoming too extreme, where the star operator's role would be to identify and potentially correct these states.

The visualization you've provided seems to be a histogram showing the distribution of values obtained from the plaquette and star operators after the simulation runs. The presence of two histograms suggests that you are comparing the distributions of two different kinds of outcomes or states in the network, likely before and after the error detection and correction algorithm is applied.

# 5. Error Syndrome Measurement Application in Social Network Models

When applying the concept of error syndrome measurement from toric codes to social network models, we can propose the following formulas and concepts for error detection and correction:

### 5.0.1 1. Representation of Network States

Each node in the network (individuals, communities, organizations) is represented by specific parameters. Indicators such as economic health, social stability, environmental impact, etc., are considered. These states are represented by a vector $\mathbf{v}$, where each element $v_i$ represents the state of node $i$.

### 5.0.2 2. Detection of Anomalies (Errors)

We define operations equivalent to the Plaquette operator and Star operator from toric codes. These operators correspond to local segments (Plaquette operator) and global segments (Star operator) of the network.

**Plaquette Operator** $B_p$: To detect local anomalies, the relationships between adjacent nodes are evaluated. For example, if the economic disparity between adjacent nodes exceeds a certain threshold, it can be detected as a local imbalance (error).

$$B_p(\mathbf{v}) = \sum_{\langle i,j \rangle} f(v_i, v_j)$$

Here, $\langle i, j \rangle$ represents pairs of adjacent nodes, and $f$ is a function measuring the relationship between the pair.

**Star Operator** $A_s$: To measure the balance of the entire system, a statistical measurement of the state of the entire network is performed. This may involve the use of overall averages, variances, or other statistical indicators.

$$A_s(\mathbf{v}) = g(\{v_i\})$$

Here, $g$ is a function assessing the state of the entire network.

### 5.0.3 1. Correction Algorithm

When an error is detected, a correction algorithm is applied. This algorithm proposes specific policies or interventions to improve the state of the network.

**Local Correction**: Targeted interventions are carried out for locally detected issues identified by the Plaquette operator. This may include resource reallocation, introduction of educational programs, infrastructure improvements, etc.

**Global Correction**: If the anomaly affects the entire system, broader policy changes may be required, as indicated by the Star operator. This could include economic policy adjustments, legal reforms, large-scale public projects, etc.

We consider the formulas and calculation processes for applying the error syndrome measurement and error correction process for a single bit-flip error in toric codes to social network models. Here, the 'bit-flip error' represents anomalies or fluctuations in social networks (e.g., economic crises, social instability), and the measurement and correction of the error syndrome are processes to detect and address these anomalies.

**Network Nodes**: Each node in the network represents individuals, communities, or organizations within a social system.

**Node State**: The state of each node is represented by a numerical value indicating the node's health, economic status, social influence, etc. A normal state is denoted as '0', and an abnormal state as '1'.

## 5.1 Error Syndrome Measurement

**Detection of Anomalies**: A sudden change in the state of a node to '1' is detected as a bit-flip error (social or economic anomaly).

**Assessment of Impact Range**: The impact of anomalies originating from a node is assessed on surrounding nodes using the adjacency matrix to consider the connections between nodes.

## 5.2 Error Correction Process

**Local Correction**: Direct interventions (such as economic or social support) are implemented for nodes where anomalies are detected.

**Global Correction**: If the anomaly affects the entire system, a broader policy review and system-level interventions become necessary.

## 5.3 Formulas

**Detection of Anomalies**:

$$E_i = \begin{cases} 1 & \text{if node } i \text{ is in an abnormal state} \\ 0 & \text{otherwise} \end{cases}$$

**Assessment of Impact Range**:

$$I_i = \sum_{j=1}^{N} A_{ij} E_j$$

Here, $I_i$ represents the total abnormal impact on node $i$, and $A_{ij}$ is the element of the adjacency matrix indicating the connection between nodes $i$ and $j$.

(1) **Anomaly Detection**: Evaluate the state of each node and identify abnormal states (bit-flip errors).

(2) **Calculation of Impact Range**: Calculate the impact range of the nodes where anomalies have occurred.

(3) **Decision of Corrective Actions**: Based on the detected anomalies and their impact range, determine appropriate interventions.

We propose formulas and calculation processes for applying the differences in the band structure of excitation energies between 'charge' (e-type) anyons and 'magnetic flux' (m-type) anyons in toric codes to a social system model. Here, the band structure represents the magnitude and range of impact that different types of 'anomalies' have on the system.

If we interpret the e-type and m-type anomalies in the context of opinion dynamics, they could represent different forms of 'opinion errors' or deviations from a norm within a community represented by the network. The e-type anomalies could represent one form of deviation, while the m-type represents another. The arrows in the visualization could indicate the 'direction' of the anomaly or deviation, with red arrows for e-type and green for m-type.

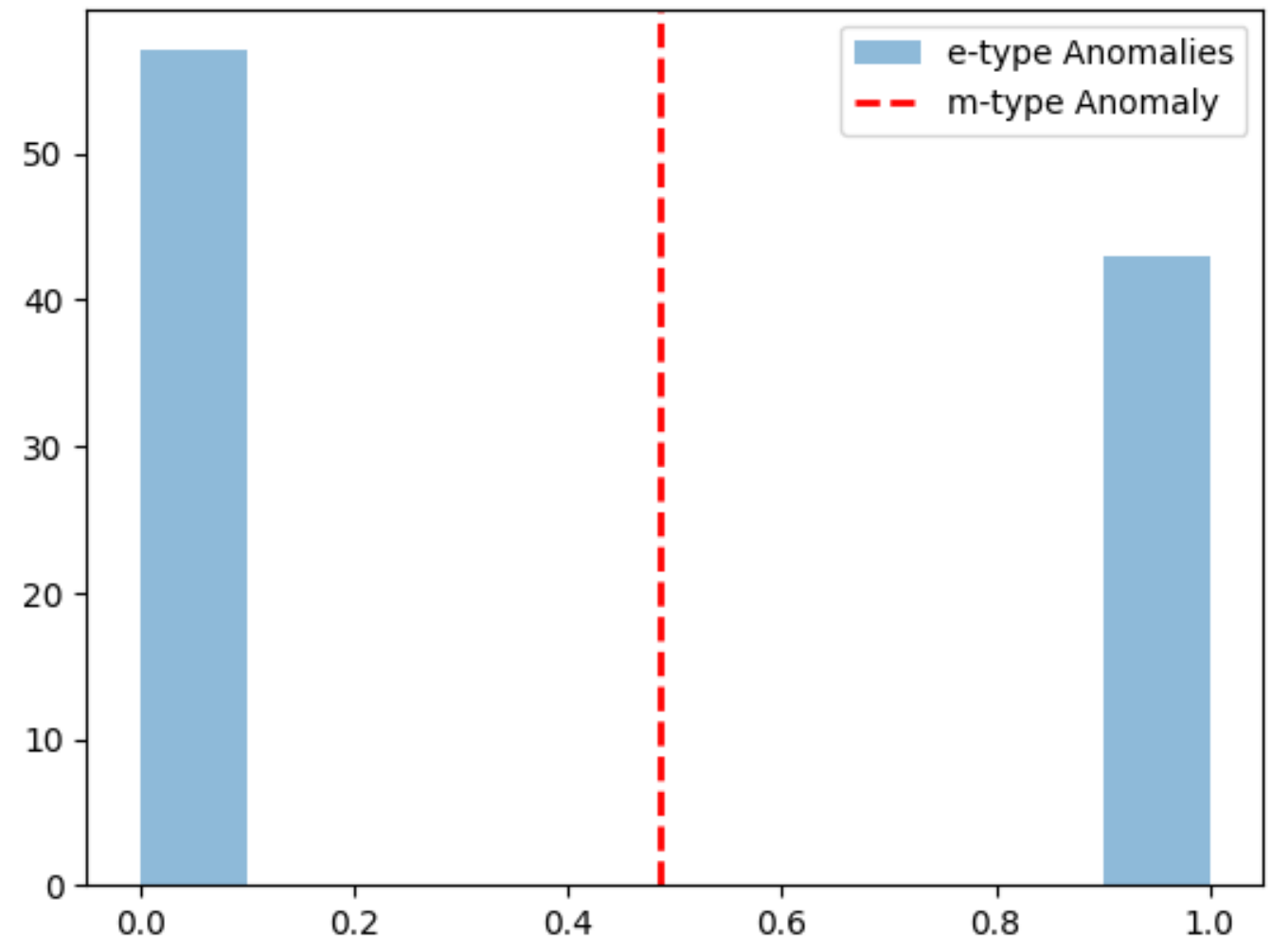

Fig. 5: Two different kinds of Anomalies

### Interactions Related to the Star Operator ($J_s$)

The star operator is not explicitly defined in the new code snippet, but typically it would check for e-type errors around a vertex in a quantum error-correcting code. In this analogy, if we were to implement a star operator, it would examine the network's nodes for patterns of e-type anomalies, potentially to correct or reduce their influence.

### Interactions Related to the Plaquette Operator ($J_p$)

Similarly, the plaquette operator would typically check for m-type errors around a face or plaquette. In the opinion dynamics analogy, it would look for m-type anomalies. Again, while not explicitly defined in your code, this operator could serve to detect and address patterns of m-type deviations in the network.

### Behavior Related to Charge (e-type) Anyons ($H_e$)

The red arrows indicate the presence of e-type anomalies. If we were to translate this into a form of energy or interaction term $H_e$, it would likely quantify the impact or 'cost' of these anomalies on the system. In a physical model, such anomalies would disrupt the ground state, while in opinion dynamics, they could represent disruptive or influential opinions that diverge from the consensus.

### Behavior Related to Flux (m-type) Anyons ($H_m$)

The green arrows represent the m-type anomalies. As with $H_e$, an interaction term $H_m$ would describe the influence of these anomalies. M-type anomalies typically interact differently than e-type anomalies in quantum models; applying this to opinion dynamics might suggest a different form of influence or disruption within the network.

You've provided shows a grid network with annotated nodes indicating the presence of e-type and m-type anomalies. This could be the starting point for a simulation where these anomalies are propagated, interact, and potentially corrected through some algorithm that mimics the error correction process in a quantum system.

To extend the analogy to opinion dynamics, one could simulate the spread of opinions (analogous to the propagation of anyons in a quantum system) and apply 'correction' mechanisms to adjust the opinions towards a desired state or consensus.

#### 5.3.1 1. e-type anyons (Charge-Type Anomalies)

e-type anyons indicate local anomalies. Their energy band structure varies based on the size and importance of the anomaly.

**Modeling**: Quantify the 'strength' or 'importance' of local anomalies. For example, indicators could be the severity of an economic crisis or the scope of social unrest.

**Formulas**:

$$E_i = f(\text{local anomaly severity at node } i)$$

Here, $f$ is a function evaluating the importance of the anomaly.

**Calculation Process**: Evaluate the importance of anomalies at each node and calculate the 'energy' of local anomalies based on this.

#### 5.3.2 2. m-type anyons (Magnetic Flux-Type Anomalies)

m-type anyons indicate global anomalies. Their energy band structure represents the magnitude of impact of anomalies that affect the entire system.

**Modeling**: Quantify the degree of overall imbalance or instability of the system. This could be an indicator of overall economic instability or widespread social dissatisfaction.

**Formulas**:

$$M = g(\text{total system imbalance})$$

Here, $g$ is a function assessing the overall imbalance of the system.

**Calculation Process**: Evaluate the impact of anomalies across the entire system and calculate the 'energy' of overall imbalance.

### 5.4 Application to Social Systems

**Local Interventions**: Plan interventions of appropriate scale and intensity based on the energy of local anomalies.

**Global Strategies**: Consider more extensive policy changes and system-level interventions based on the energy of global anomalies.

This approach allows for more effective identification of different types of problems and optimal responses to each. By using the concept of band structure, the 'weight' and 'impact' of problems can be understood in more detail, allowing for the prioritization of interventions based on these aspects.

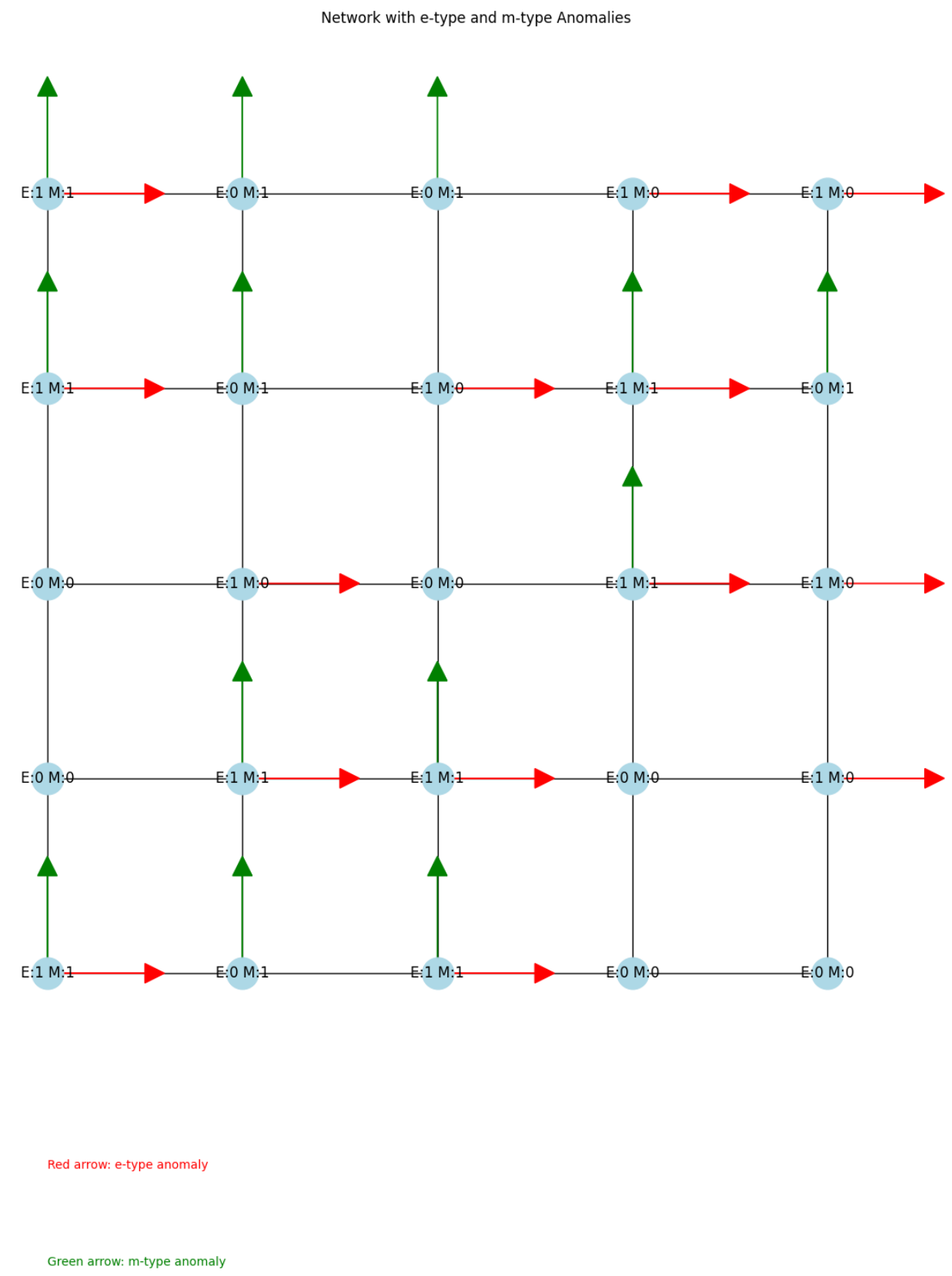

Fig. 6: Network with e-type and m-type Anomalies

### Opinion Dynamics Consideration

- In the context of opinion dynamics, the network could represent individuals (nodes) in a social network, and the arrows (anomalies) could represent differing opinions or influences. The red arrows (e-type anomalies) could signify one type of influence or opinion, while the green arrows (m-type anomalies) could signify another. This depiction might be used to visualize the distribution of these opinions or influences and to understand how they might interact.

## Interaction Related to the Star Operator ($J_s$)

- The star operator in quantum error correction is associated with the vertices of a lattice and is used to detect errors. In this model, if we were to apply such an operator, it could represent a form of local consensus mechanism, where the red arrows (e-type anomalies) indicate points of disagreement that need to be addressed.

## Interaction Related to the Plaquette Operator ($J_p$)

- The plaquette operator usually applies to the faces of a lattice and detects different types of errors. In this analogy, the green arrows (m-type anomalies) might represent another form of opinion or influence that is spread over a group of individuals. The plaquette operator could be seen as a mechanism for identifying and resolving broader consensus issues within the network.

## Behavior Related to Charge (e-type) Anyons ($H_e$)

- In a physical system, e-type anyons represent errors that can be detected by the plaquette operator. If we translate this to the network model, it could suggest areas where individual opinions (e-type anomalies) are in conflict with the surrounding consensus, which might require specific strategies to resolve.

## Behavior Related to Flux (m-type) Anyons ($H_m$)

- M-type anyons in a physical system are errors that can be detected by the star operator. In the social network model, this might point to regions where the overall opinion or influence (m-type anomalies) is in conflict with the local individual opinions, suggesting a need for broader strategies that address group dynamics.

The illustration of the network with labeled anomalies can be used to simulate how these opinions or influences might propagate, interact, and possibly be corrected over time. Such simulations could be valuable in studying complex social dynamics, understanding the spread of information, and managing social consensus or conflict.

### 5.4.1 e-type Values

## Opinion Dynamics Consideration

- In terms of opinion dynamics, the nodes could represent individuals in a social network, and the S and P values could represent two different aspects of their opinions, such as strength and spread, respectively. The varying colors in the nodes reflect the diversity of these aspects within the population.

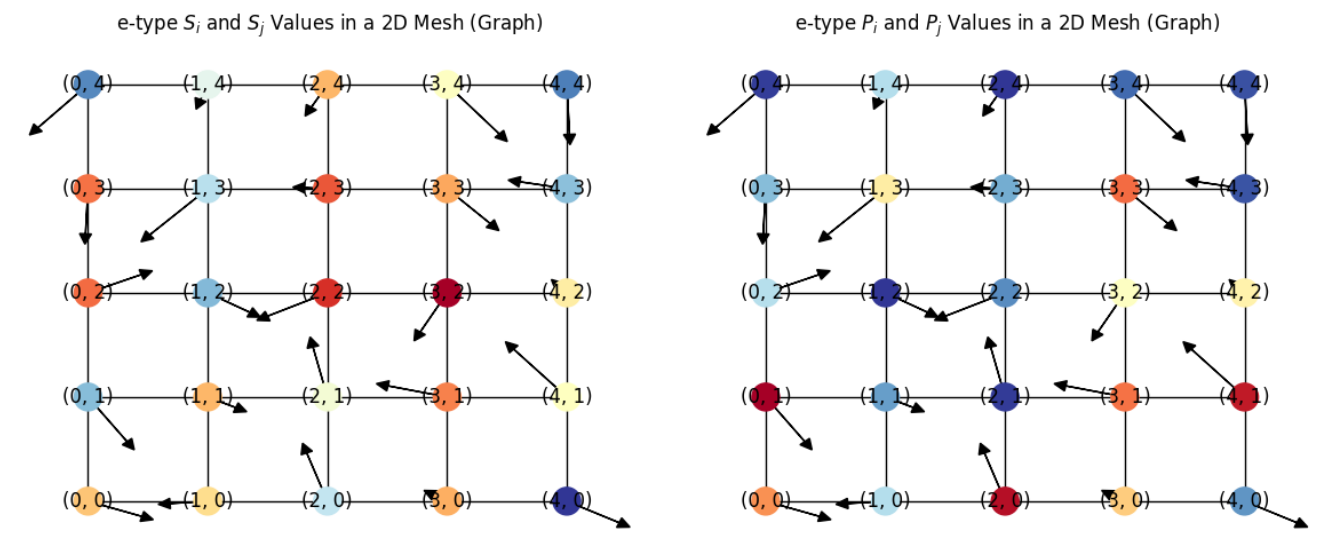

Fig. 7: e-type Values in a 2D Mesh

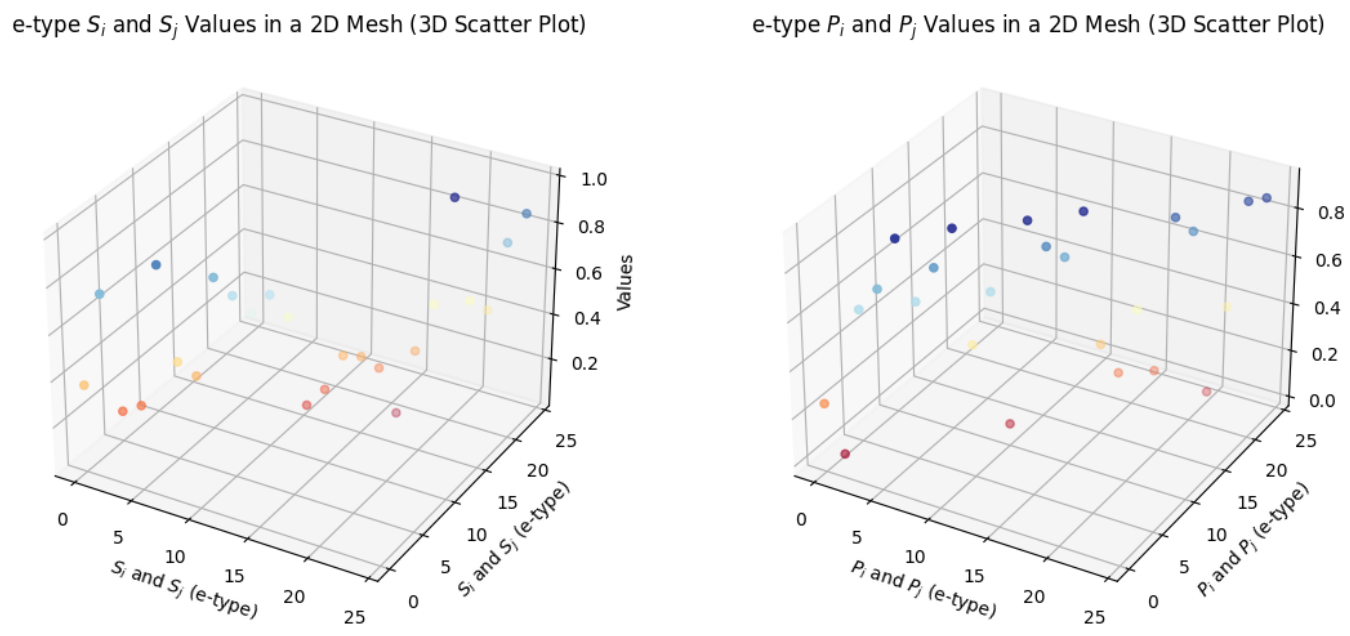

Fig. 8: e-type Values in a 3D Mesh

## Interactions Related to the Star Operator ($J_s$)

- The star operator in a quantum error correction context is used to detect and correct errors around a vertex. Translated to opinion dynamics, $J_s$ could represent the influence of an individual's opinion based on the opinions of their immediate neighbors, suggesting a local consensus mechanism.

## Interactions Related to the Plaquette Operator ($J_p$)

- The plaquette operator usually applies to a plaquette (a square or face of the grid) in error correction codes and detects errors within that plaquette. In the analogy, $J_p$ could represent the influence of a group's collective opinion on the individual, pointing to a broader societal influence mechanism.

## Behavior Related to Charge (e-type) Anyons ($H_e$)

- In the physical system, e-type anyons are associated with errors on qubits that the plaquette operator can detect. In the social model, this might correlate with discrepancies or 'errors' in local opinion consensus, where e-type values indicate individual deviations from a local norm.

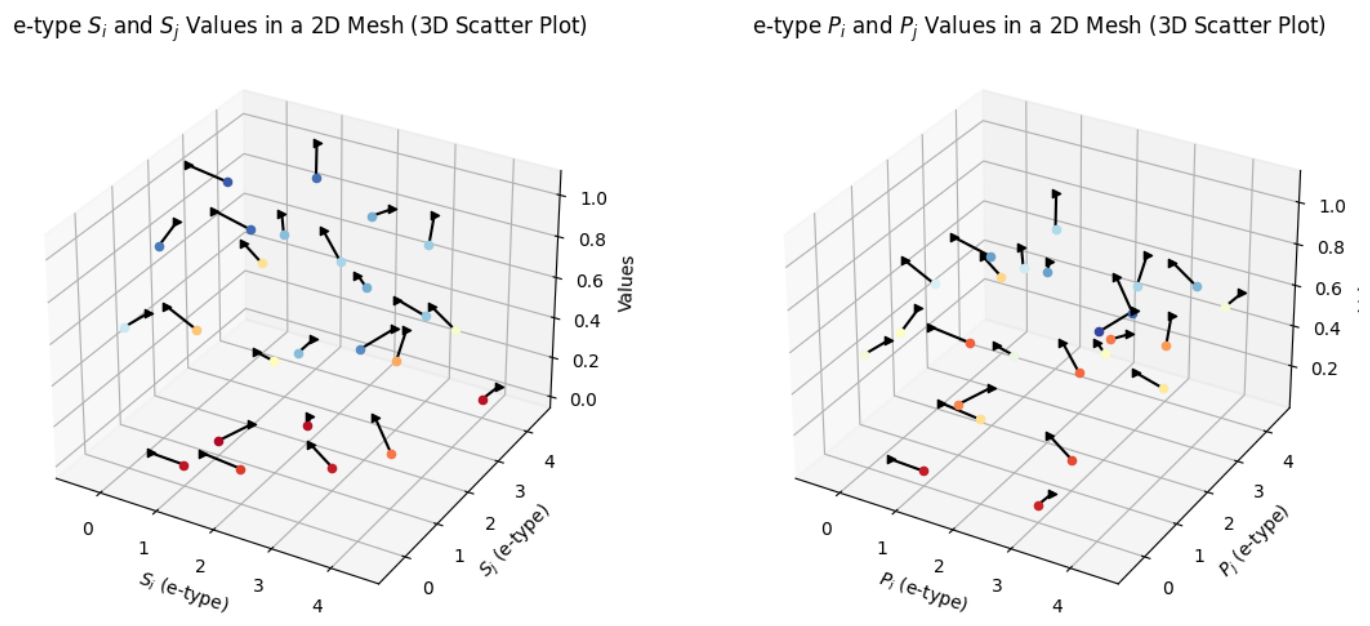

Fig. 9: e-type Values in a 3D Mesh

## Behavior Related to Flux (m-type) Anyons ($H_m$)

- Similarly, m-type anyons are associated with errors that the star operator can detect. In the social network model, this could relate to individuals whose opinions are becoming too extreme or isolated from the broader consensus, indicated by m-type values.

The arrows depicted in the images may symbolize the tendency or movement of opinion changes in the network, with red arrows possibly signifying a shift towards one opinion and green arrows a shift towards another. The 3D scatter plots further visualize how these opinions (or errors, in quantum terms) are distributed across the network, potentially allowing us to observe clusters, outliers, or patterns in opinion dynamics.

In summary, the simulation captures the complexity of opinion formation and change, influenced by local interactions ($J_s$) and global societal trends ($J_p$), with the individual nodes possibly representing agents undergoing changes in opinion (anomalies) over time. This model could be insightful for understanding the spread of information, the formation of consensus, or polarization within a community.

### 5.4.2 m-type Values

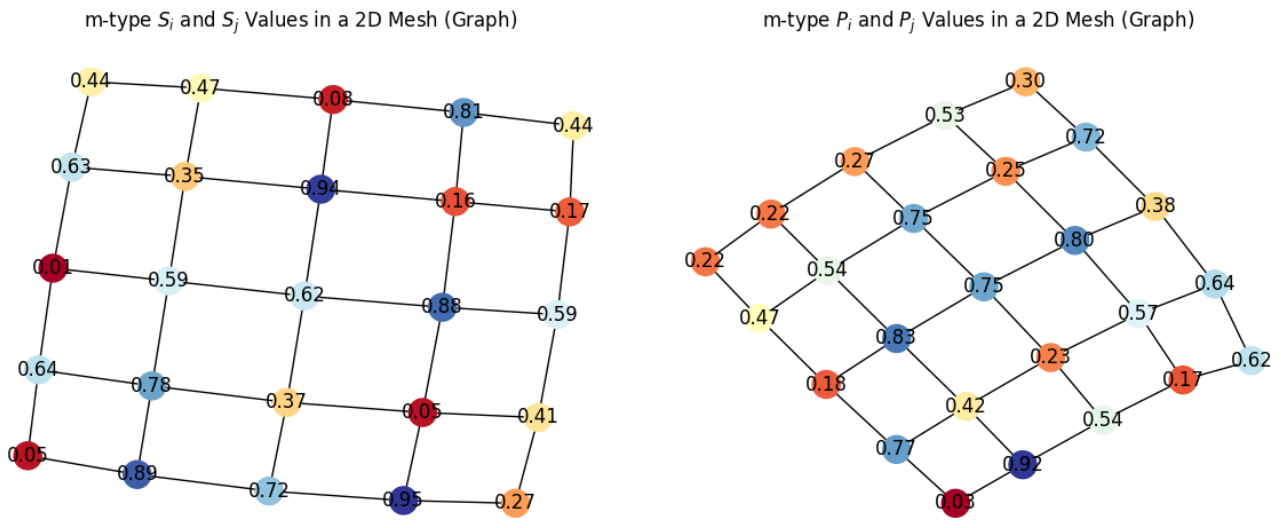

Fig. 10: m-type Values in a 2D Mesh

## Opinion Dynamics Consideration

If we consider this model as a representation of opinion dynamics, the nodes in the 2D mesh can be seen as individuals in a social network. The S and P values could signify different aspects of their opinions, such as certainty and persuasiveness. The different colors could indicate the strength or direction of these opinions. The network shows the diversity and distribution of these opinions among the individuals.

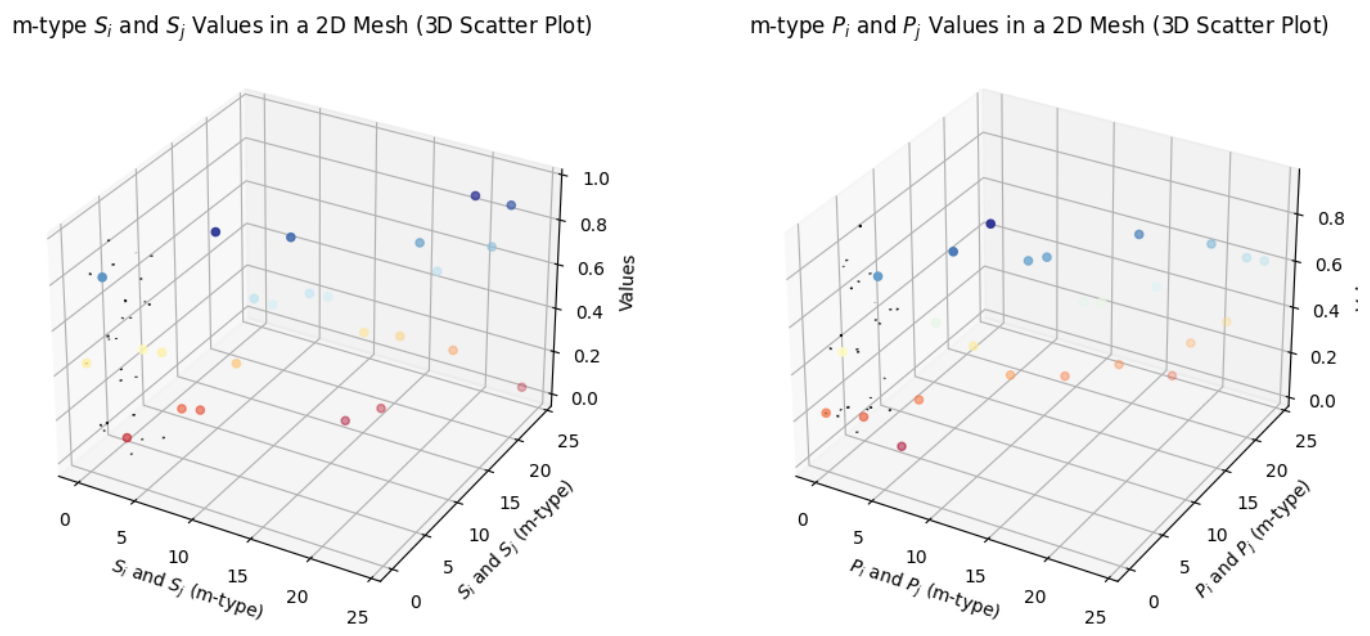

Fig. 11: m-type Values in a 3D Mesh

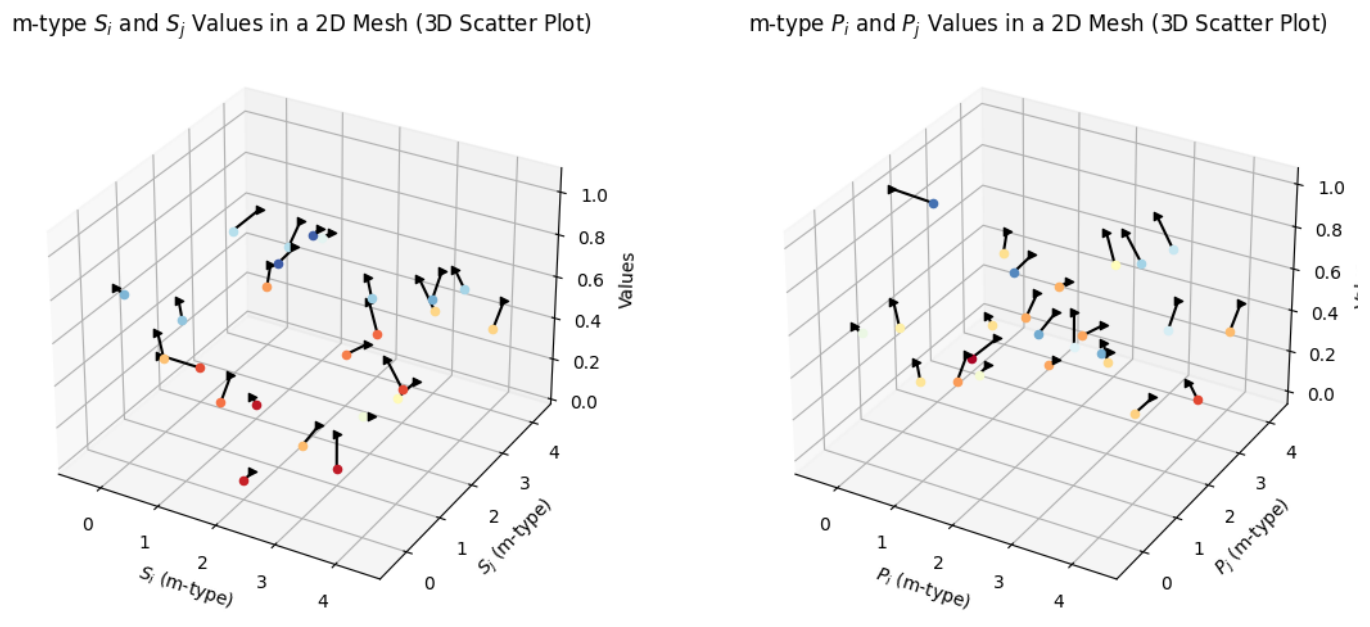

Fig. 12: m-type Values in a 3D Mesh

## Interactions Related to the Star Operator ($J_s$)

- In the context of quantum error correction, the star operator applies to vertices and is used to detect certain types of errors. In an opinion dynamics model, $J_s couldrepresentasocialinfluencethataltersth eopinionsofindivic$

## Interactions Related to the Plaquette Operator ($J_p$)

- $J_p$, related to the plaquette operator, could be analogous to the broader societal norms that impact individual opinions in a social network model. This operator might detect and correct for deviations from the broader societal consensus.

## Behavior Related to Charge (e-type) Anyons ($H_e$)

- In a quantum system, $H_e$ would be the term in the Hamiltonian related to the e-type anyons, which could be thought of as errors or charges in the system. Translated into opinion dynamics, this might reflect the presence of a minority opinion or a dissenting individual in a community that is otherwise in agreement.

ion is divided.

# 6. Conclusion

### 6.0.1 Ising Model to Social Simulation

When applying the Ising model to social simulations, it is crucial to correlate each element of the model with phenomena and concepts in social systems. Below, we propose interpretations of formulas and parameters for the application of the Ising model to social simulations.

#### Spin States

**Spin State** $s_i$: Represents the opinions or states of individual agents (individuals, groups). +1 can denote adopting a specific opinion or behavior, while -1 represents the opposite.

#### Energy Function

**Interaction** $J_{ij}$: Represents the strength of influence or relationships between agents. A higher value implies stronger interaction, modeling the degree of social connections or influence.

**External Field** $h$: Represents external influences or pressures. This can indicate factors like media influence, government policies, social trends, etc., affecting the opinions or states of individual agents.

### 6.0.2 Bit-Flip Error (Opinion Change)

**Bit-Flip**: Represents a sudden change in an agent's opinion or behavior. This can occur due to new information or external influences.

### 6.0.3 Energy Change and Error Correction

**Energy Change** $\Delta E$: Indicates an increase in anomalies or instability in the social system. An increase in energy is interpreted as disrupting social harmony or stability.

**Error Correction**: Represents mechanisms for restoring stability in the social system. This could be social mechanisms promoting opinion harmony or processes for individuals to reassess their opinions.

**State Change of Agents**:

$$s_i \rightarrow -s_i$$

**Energy of the Social System**:

$$E = -\sum_{\langle i,j \rangle} J_{ij} s_i s_j - h \sum_i s_i$$

**Energy Change Due to State Change**:

$$\Delta E = E_{\text{after}} - E_{\text{before}}$$

(1) **Setting the Initial State**: Set the initial opinions or states of each agent in the social system.

(2) **Energy Calculation**: Calculate the total energy of the social system in its initial state.

(3) **Introduction of Opinion Change**: Invert the opinion of a randomly selected agent.

(4) **Assessment of Energy Change**: Calculate the energy change in the social system due to the opinion change.

(5) **Error Correction**: If energy increases, revert the agent's opinion to its original state or adopt a new state.

This approach allows for simulating the dynamics of opinion formation and collective behavior in social systems and exploring strategies for maintaining social harmony and stability.

## 6.1 Agent States

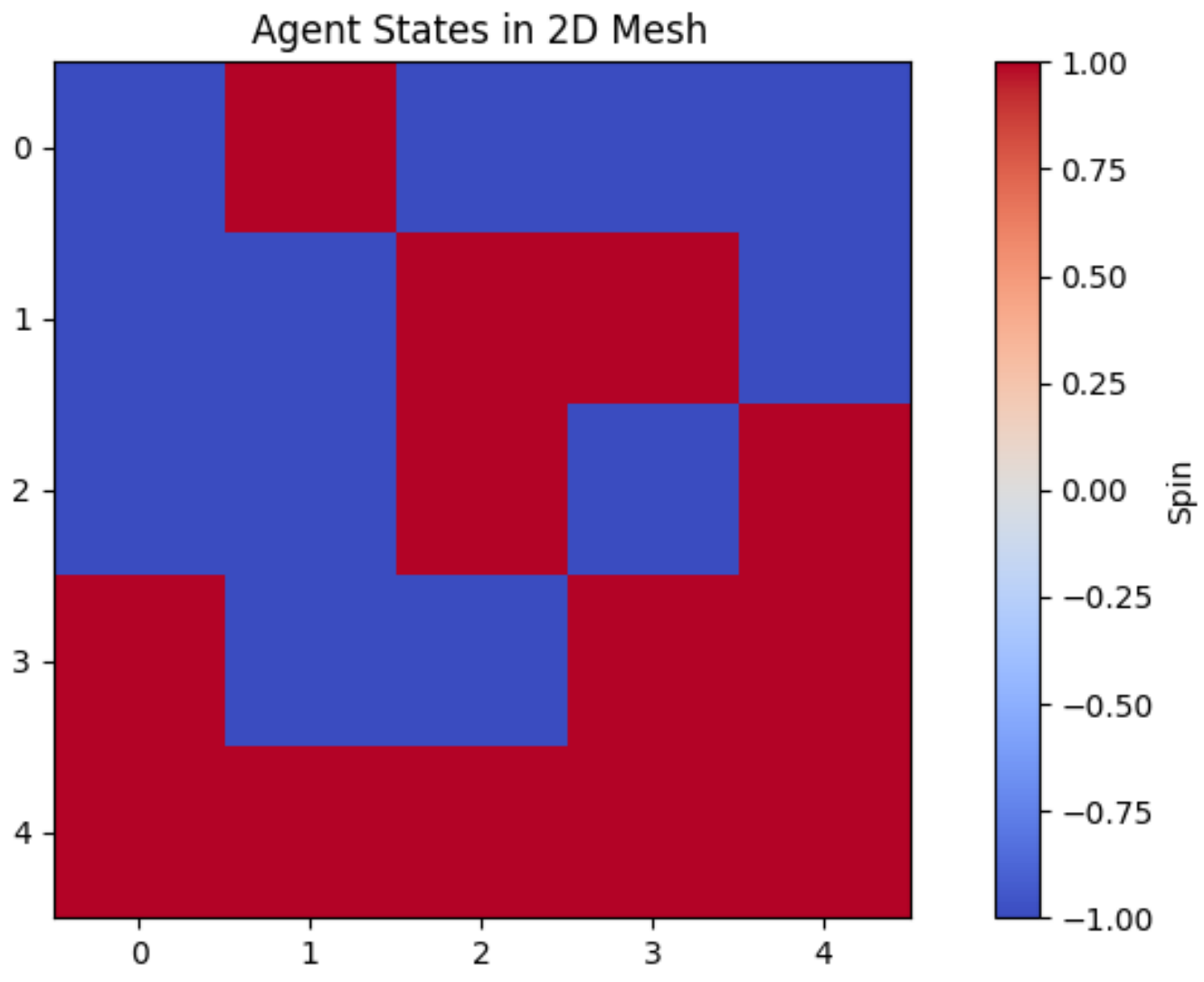

Fig. 13: Agent States in 2D Mesh

## Opinion Dynamics Consideration

The 2D and 3D mesh plots represent a distribution of opinions across a network of agents. The "spins" could symbolize agreement (+1, often represented by blue) or disagreement (-1, often represented by red) with a particular stance. The dynamics would then involve how these opinions change over time, influenced by their neighbors (modeled by the interaction coefficients $J$) and an external factor (modeled by the external field $h$). The tendency of spins to align or oppose each other can be influenced by the interaction coefficients $J$, which represent the strength and sign of the pairwise influence between agents. Positive $J$ values encourage neighboring spins to align, while negative $J$ values encourage opposition. The external field $h$ can represent external media influence or societal pressure that biases the agents' spins in a certain direction. In the context of opinion dynamics, an "error" could represent an agent holding a dissenting opinion in a local consensus. The simulation allows for spontaneous "errors" where an agent's spin (opinion) flips, which can represent a change of opinion due to personal reflection or external influences.

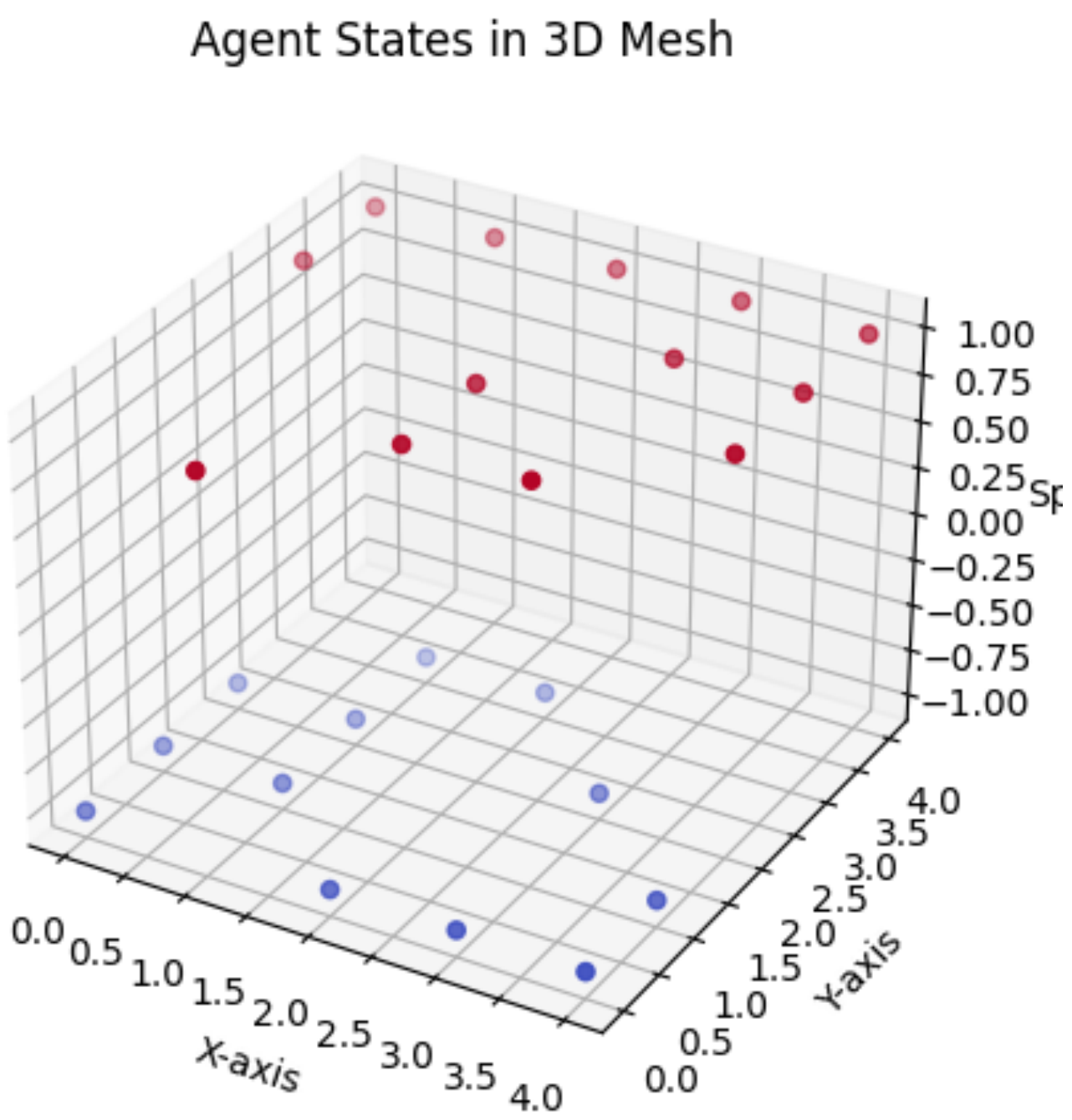

Fig. 14: Agent States in 3D Mesh

### Tendency for Bit-flips

A bit-flip in this model is a change in the state of an agent's spin. This could be analogous to an individual changing their opinion from agree to disagree or vice versa. The model as currently constructed allows for random bit-flips, but in a more sophisticated model, this could be made dependent on the energy difference ($\Delta E$) induced by the flip, with less likely flips occurring when $\Delta E$ is positive and more likely when $\Delta E$ is negative.

This kind of simulation can be particularly valuable for studying phenomena such as social conformity, the spread of misinformation, and the polarization of opinions. It can help in understanding how local interactions and external pressures contribute to the overall opinion landscape of a society or a social network.

## 6.2 Transversal Gates and Gauge Fixing Strategies (e-type)

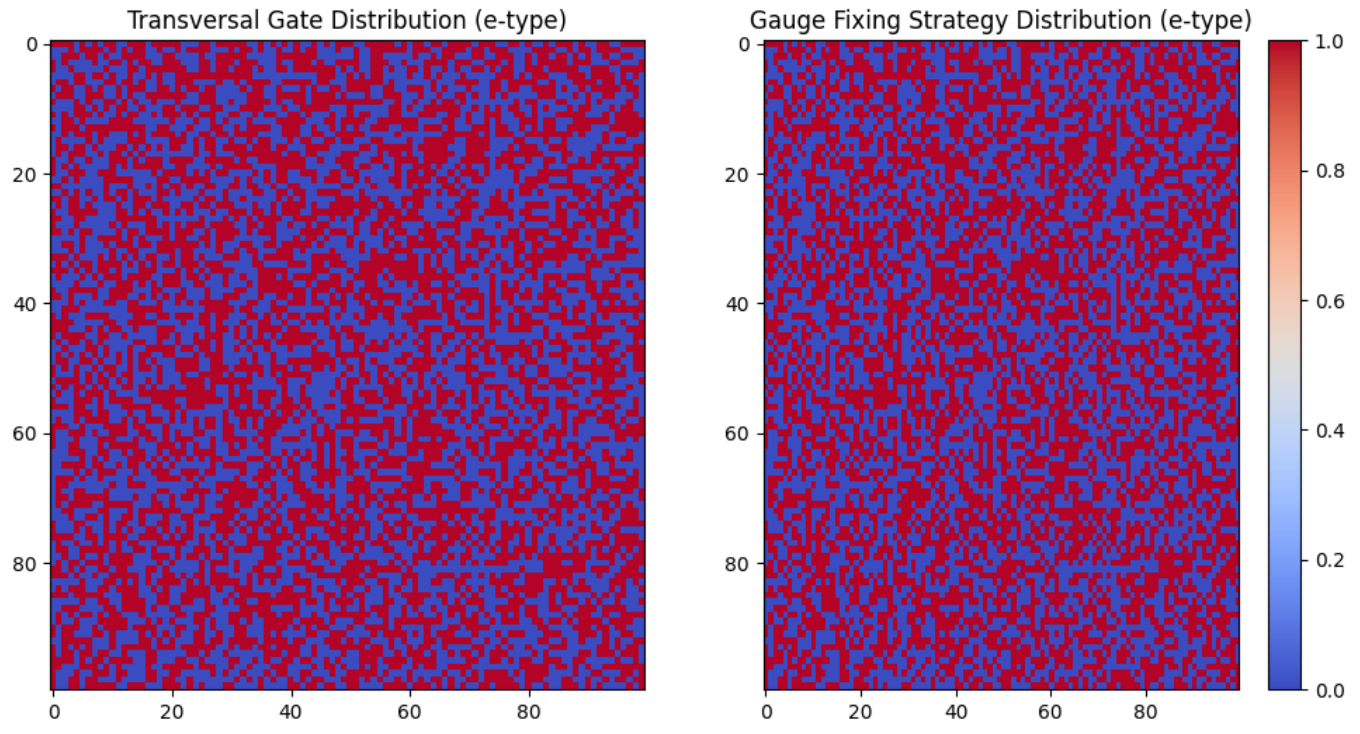

Fig. 15: Transversal Gate Distribution and Gauge Fixing Strategy Distribution(e-type)

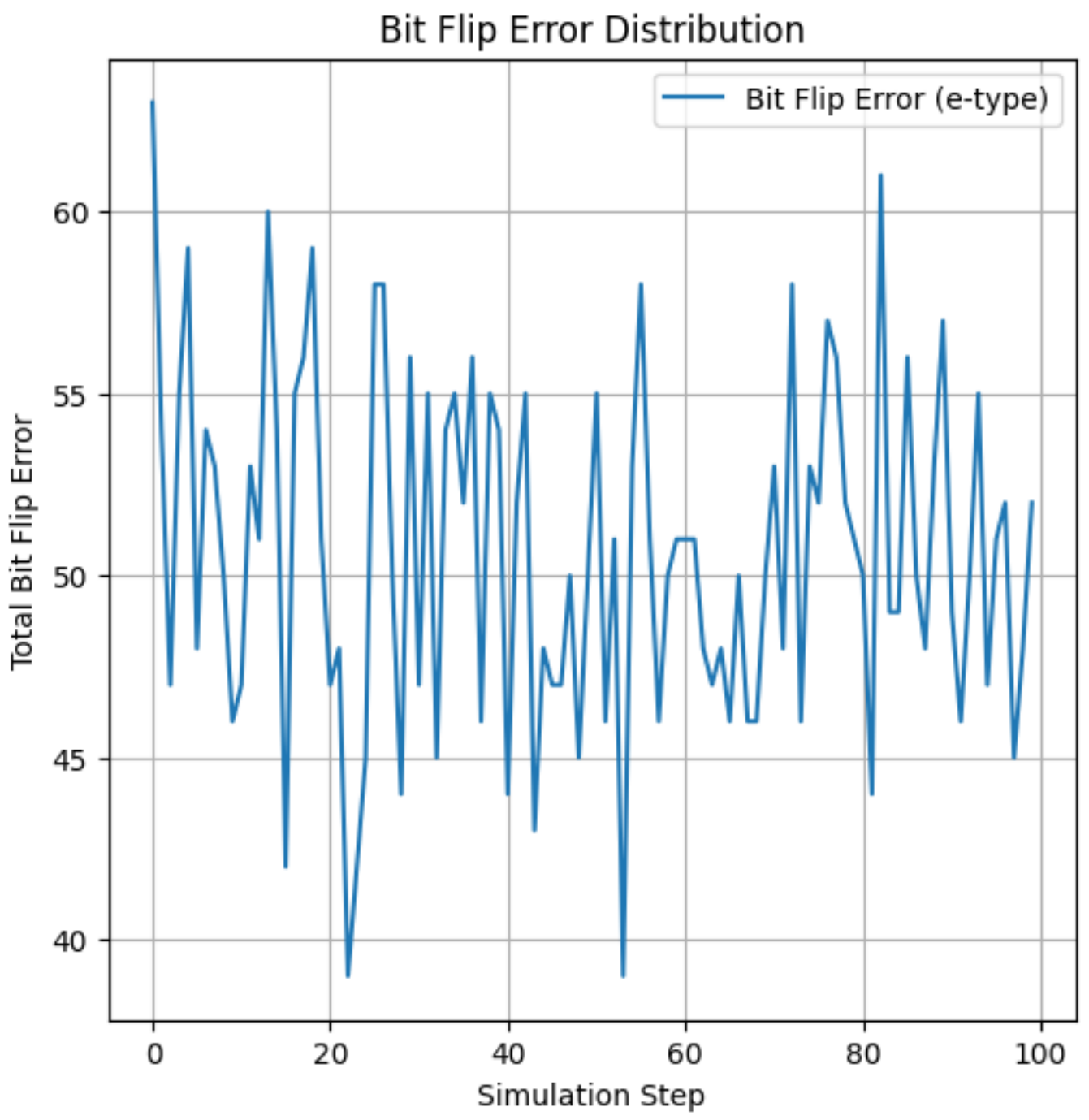

Fig. 16: Total Bit Flip Error(e-type)

### Opinion Dynamics Consideration

The visualizations can represent a society's distribution of opinions, where each individual (or agent) holds a binary stance on an issue (e.g., agree/disagree, yes/no). The "Transversal Gate Distribution" and "Gauge Fixing Strategy Distribution" could reflect different strategies for reaching consensus or correcting societal 'errors' — where opinions differ from the expected norm.The simulation could model how opinions (or errors, in a quantum context) spread and are corrected within a population. The heatmaps might indicate the prevalence of certain opinions after a widespread event or policy (transversal gate) and the subsequent attempts at social regulation or norm enforcement (gauge fixing).

## Behavior Related to Charge (e-type) Anyons ($H_e$)

In the context of this simulation, $H_e$ might represent the 'energy' or 'cost' associated with the distribution of these opinions or stances within the society. A more uniform distribution (all blue or all red) could correspond to a lower 'energy' state, reflecting a consensus or common opinion, while a mixed distribution indicates higher 'energy', reflecting societal disagreement or conflict. In terms of quantum physics, H$_e would be related to the presence of e - type anyons, which could be interpreted as errors or deviations fr$ ... $l behav$

## Tendency of Transversal Gates

The transversal gate typically applies a corrective operation across a system. In the simulation, this could correspond to a societal mechanism that attempts to correct or flip the opinion of a randomly chosen individual, aiming to reduce the overall 'error' or disagreement.

## Trend of Gauge Color Codes

Gauge color codes are typically used in quantum error correction to detect and correct errors. In this model, they might represent an underlying belief or value system that influences opinion formation. The tendency here would be how these systems are upheld or changed over time, potentially as a response to societal pressures or changes. The trend in the gauge color codes could suggest how social norms and opinions are influenced over time, with potential shifts towards or away from a consensus.

## Tendency of Gauge Fixing Strategies

A gauge fixing strategy in this context could represent the societal norms or laws that 'fix' the opinions or behaviors of individuals to align with a certain standard. The distribution reflects the effectiveness and patterns of these strategies over the society.

## Tendency for Bit Flips

Bit flips in a social model could represent individuals changing their opinions. The "Bit Flip Error Distribution" graph shows the total number of opinion changes over time, reflecting the volatility or stability of opinions within the society.

## The "Transversal Gate Distribution" and "Gauge Fixing

Strategy Distribution" heatmaps show the state of each individual's opinion at each simulation step, while the line graph of "Bit Flip Error Distribution" over simulation steps provides insight into the dynamics of how these opinions change over time. Peaks in the graph could indicate times of high societal tension or significant events causing many individuals to change their opinions.

## 6.3 Transversal Gates and Gauge Fixing Strategies (m-type)

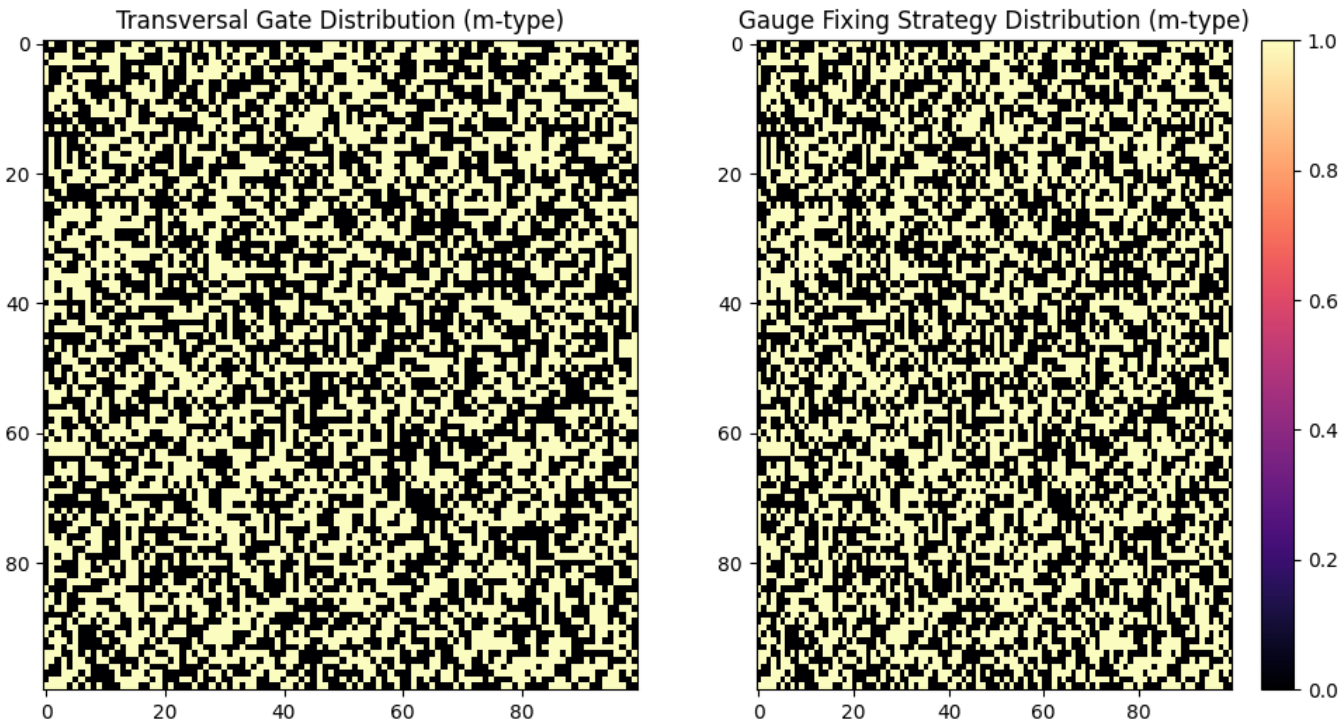

Fig. 17: Transversal Gate Distribution and Gauge Fixing Strategy Distribution (m-type)

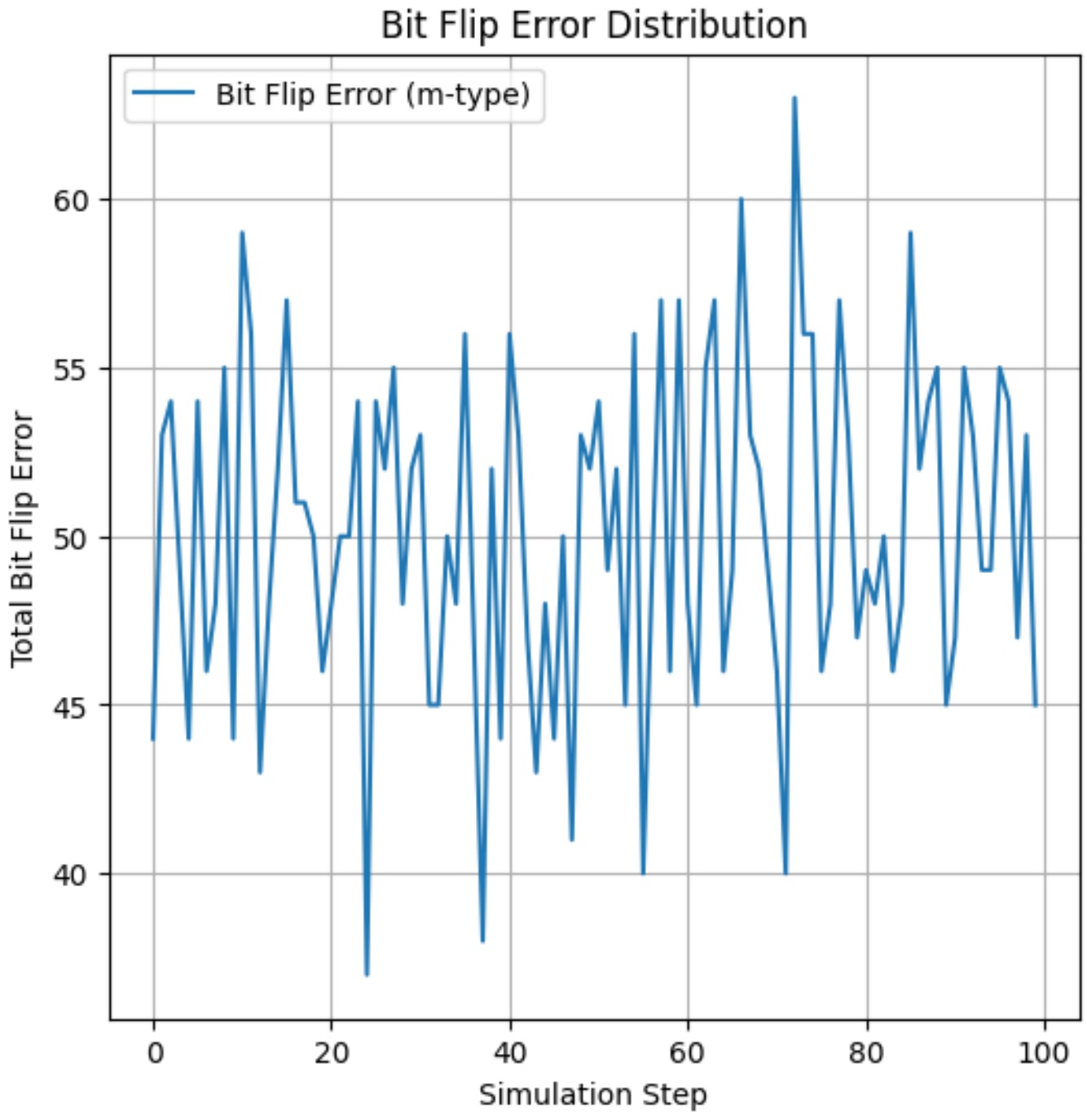

Fig. 18: Total Bit Flip Error(m-type)

## Opinion Dynamics Consideration

In the realm of opinion dynamics, m-type entities could represent a different dimension or type of opinion compared to e-type entities. The distribution of these opinions across a population or network could be visualized through the "Transversal Gate Distribution" and "Gauge Fixing Strategy Distribution" images. These visualizations could show the diversity of opinions and how they might be influenced by corrective measures or strategies to achieve consensus.

## Behavior Related to Flux (m-type) Anyons ($H_m$)

In a system analogous to magnetic flux behavior, $H_m$ could represent the 'energy' or 'tension' due to the distribution of m-type entities across the lattice. A high variance in the distribution (both black and yellow regions in the heatmap) could indicate areas of high 'energy', which might correspond to conflicts or areas where opinions are highly polarized.

## Tendency of Transversal Gates

Transversal gates typically apply a correcting operation across an entire system. In the context of the simulation, they could be seen as a broad strategy to change individual states in an attempt to correct or align them with a certain pattern, mimicking the correction of errors in quantum error correction or the alignment of societal opinions.

## Trend of Gauge Color Codes

Gauge color codes in quantum computing are used to detect and correct errors within a system. In the social dynamics analogy, these could represent the underlying cultural or societal norms that dictate the acceptable range of opinions or behaviors. The trend would be indicative of how societal norms are maintained or evolve over time.

## Tendency of Gauge Fixing Strategies

Gauge fixing strategies in this simulation might represent the mechanisms by which a society enforces conformity or manages diversity. The distribution reflects how effective these strategies are in creating uniformity or allowing diversity within the population's opinions.

## Tendency for Bit Flips

A bit flip in a social model could represent a change in individual opinion. The "Bit Flip Error Distribution" graph indicates how often these opinion changes occur over time. Fluctuations in the graph could reflect the societal response to external events, indicating times of instability or widespread changes in opinion.

## The "Transversal Gate Distribution" and "Gauge Fixing

Strategy Distribution" heatmaps depict the state of the system at different simulation steps, possibly before and after the application of strategies or corrective measures. The line graph represents the total number of opinion changes (bit flips) at each step, providing insight into the dynamics of opinion change over time.

This type of simulation and analysis can be useful for studying the effects of different policies or strategies on public opinion, understanding the spread of information in social networks, and predicting the stability of societal consensus.

## 6.4 Transversal Gates and Gauge Fixing Strategies (Opinion Change)

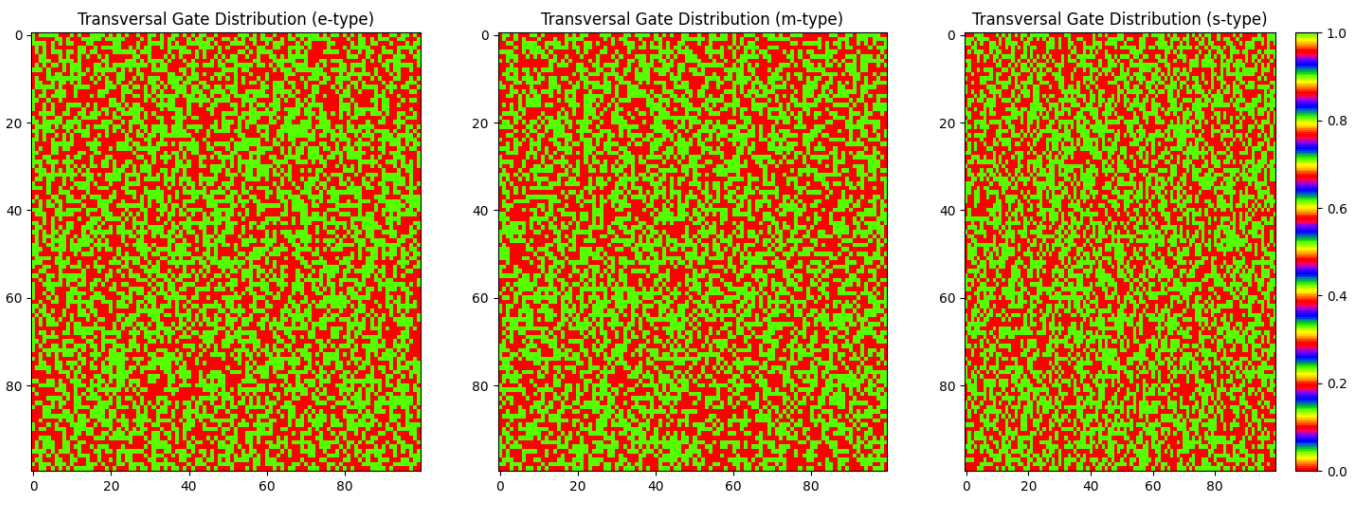

Fig. 19: Transversal Gate Distribution

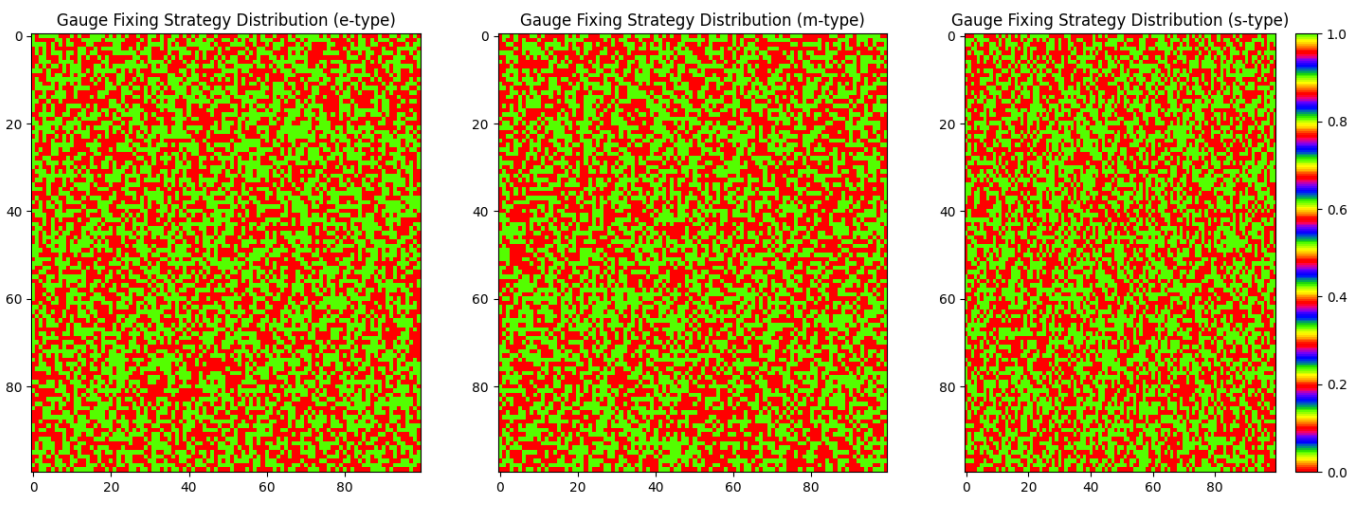

Fig. 20: Gauge Fixing Strategy Distribution

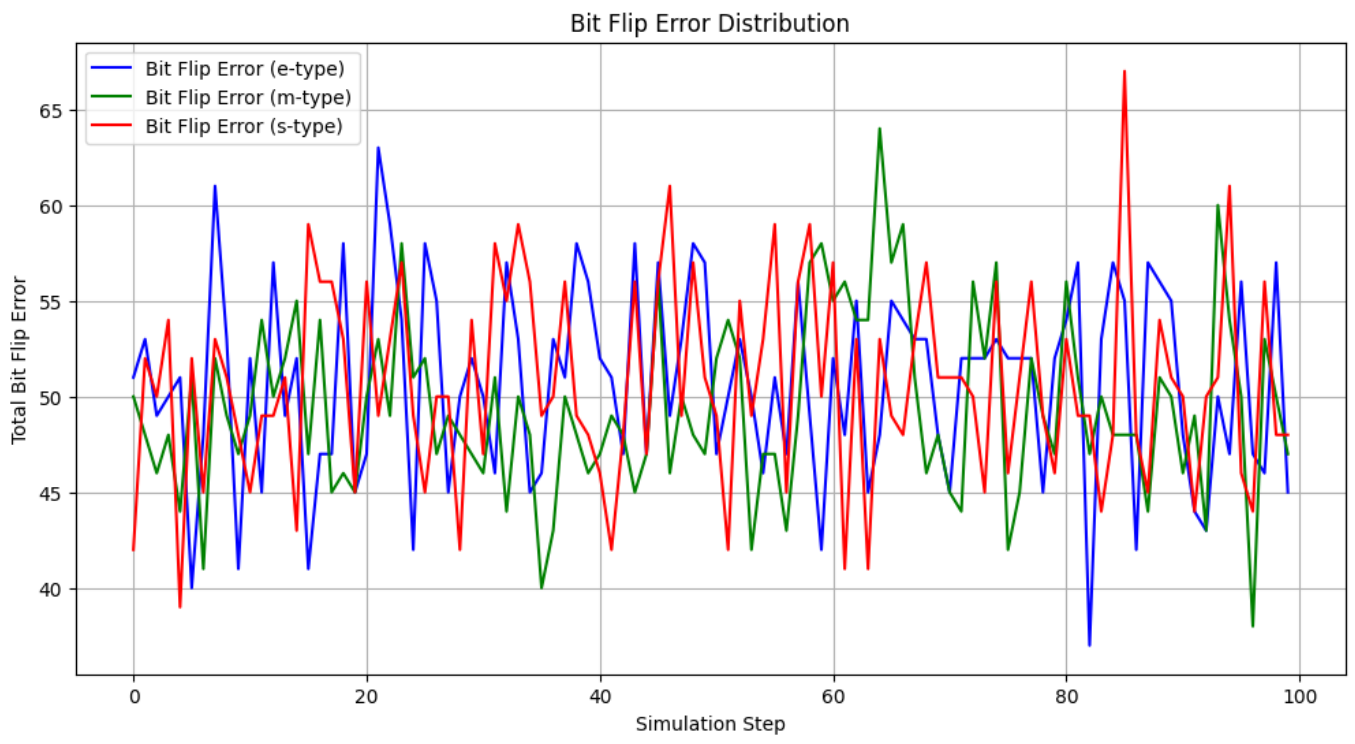

Fig. 21: Total Bit Flip Error

## Opinion Dynamics Consideration

Opinion dynamics models study how individual opinions evolve within a network and can be influenced by factors such as social influence, connectivity, and personal propensity. In the context of the model, the 'opinions' may represent the state of the nodes (whether they are in an error state or not). The random initialization and subsequent modifications of the states can be seen as a simplified representation of opinion dynamics where each node's state is influenced randomly rather than by its neighbors.In this case, each node has a binary state that can flip based on random processes or through the application of gates and strategies, analogous to how an individual's opinion might change due to external influences or internal decision-making.

## $H_e$ - Charge (e-type) Anyons

$H_e$ seems to represent a Hamiltonian component related to e-type anyons, which are abstractions used in topological quantum computing to represent quasiparticles that can arise in a two-dimensional system. The behavior of these anyons within the region could relate to how errors (represented by these anyons) are distributed and corrected over time. The visualization suggests that errors are randomly distributed and corrected at each step, possibly simulating the random appearance of errors and their correction in a quantum system.The first of the heatmap images shows the e-type transversal gate distribution, and the third heatmap shows the e-type gauge fixing strategy distribution. Both heatmaps indicate a random distribution of the e-type gauge color codes over time. The behavior of these e-type anyons would reflect the random introduction of errors and their correction in a simulated quantum system. Since the distribution appears random at each step, it suggests that the error correction strategy does not have a consistent or deterministic pattern in addressing these errors.

## $H_m$- Magnetic Flux (m-type) Anyons

Similarly, $H_m$ likely corresponds to a component of the Hamiltonian related to m-type anyons, which might represent another type of quasiparticle or error state. The behavior within the region would also be indicative of the distribution and correction of these types of errors. The fact that both e-type and m-type anyons are being considered suggests that the model could be simulating a topological quantum system that can experience two distinct types of localized errors.The behavior of m-type anyons, reflected in the second heatmap for transversal gates and the middle heatmap for gauge fixing strategy, shows a similar pattern to the e-type. The m-type gauge codes also seem to be randomly distributed over time, which again could suggest the stochastic nature of error introduction and correction. The lack of a clear pattern could either be intentional (simulating random environmental noise) or it could indicate an area for improvement in the error correction strategy.

## Gauge Fixing Strategy Trends

The transversal gate distributions for e-type, m-type, and s-type (as seen in the first image) are shown as heatmaps. The randomness in the color variation indicates that the transversal gates are applied in a non-uniform, random manner across different simulation steps. This could suggest a model where errors are corrected without a specific pattern, perhaps reflecting the unpredictability of error occurrence in quantum systems.

Gauge fixing is a procedure to reduce the degrees of freedom in a gauge theory. In the simulation, the gauge fixing strategy is applied after the transversal gates, which could imply a two-step error correction process where errors are first attempted to be corrected locally (transversal gates) and then globally (gauge fixing).

## Bit Flip Error Trends

The line graph showing bit flip error distribution indicates that the total number of errors for each anyon type fluctuates over time, with no clear trend towards reduction or stabilization. This could imply that while the error correction methods (transversal gates and gauge fixing strategies) are active, they may not be optimally reducing the total number of errors. The fluctuations could also represent the balance between error introduction and correction.

In conclusion, the simulation appears to model a quantum error correction scenario with stochastic error generation and correction. The randomness in the transversal gate and gauge fixing strategy distributions, as well as the fluctuations in the bit flip error counts, suggest that the system is under constant change with no stable state being reached within the simulation steps observed. This could be representative of a noisy quantum environment where error correction is an ongoing challenge, and the strategies used do not converge to a fault-tolerant state but rather attempt to manage errors as they occur.

# Aknowlegement

The author is grateful for discussion with Prof. Serge Galam and Prof.Akira Ishii. This research is supported by Grant-in-Aid for Scientific Research Project FY 2019-2021, Research Project/Area No. 19K04881, "Construction of a new theory of opinion dynamics that can describe the real picture of society by introducing trust and distrust".

# References

zh